\documentstyle[12pt]{article}
\begin{document}

\baselineskip=0.7cm
\newcommand{\EQ}{\begin{equation}}
\newcommand{\EN}{\end{equation}}
\newcommand{\EQA}{\begin{eqnarray}}
\newcommand{\EQN}{\end{eqnarray}}
\newcommand{\EQAN}{\begin{eqnarray*}}
\newcommand{\EQNN}{\end{eqnarray*}}

\newcommand{\e}{{\rm e}}
\newcommand{\Sp}{{\rm Sp}}
\renewcommand{\theequation}{\arabic{section}.\arabic{equation}}
\newcommand{\Tr}{{\rm Tr}}
\renewcommand{\thesection}{\arabic{section}.}
\renewcommand{\thesubsection}{\arabic{section}.\arabic{subsection}}
\makeatletter
\def\section{\@startsection{section}{1}{\z@}{-3.5ex plus -1ex minus 
 -.2ex}{2.3ex plus .2ex}{\large}} 
\def\subsection{\@startsection{subsection}{2}{\z@}{-3.25ex plus -1ex minus 
 -.2ex}{1.5ex plus .2ex}{\normalsize\it}}
\def\appendix{
\par
\setcounter{section}{0}
\setcounter{subsection}{0}
\def\thesection{\Alph{section}}}
\makeatother
\def\thefootnote{\fnsymbol{footnote}}
\begin{flushright}
hep-th/9907029\\
UT-KOMABA/99-8\\
July 1999
\end{flushright}
\vspace{1cm}
\begin{center}
\Large
Generalized AdS-CFT  Correspondence \\for Matrix Theory 
in the Large N limit

\vspace{1cm}
\normalsize
{\sc Yasuhiro Sekino}
\footnote{
e-mail address:\ \ sekino@hep1.c.u-tokyo.ac.jp}
and 
{\sc Tamiaki Yoneya}
\footnote{
e-mail address:\ \ tam@hep1.c.u-tokyo.ac.jp}
\\
\vspace{0.3cm}
 {\it Institute of Physics\\
University of Tokyo, Komaba, Tokyo 153 }

\vspace{1.3cm}

Abstract\\

\end{center}
\noindent
Guided by the generalized conformal symmetry, 
we investigate the extension of AdS-CFT correspondence 
to the matrix model of D-particles in the large $N$ limit. 
We perform a complete harmonic analysis of the 
bosonic linearized fluctuations around a heavy D-particle 
background in IIA supergravity in 10 dimensions 
and find that the 
spectrum precisely agrees with that of the 
physical operators of Matrix theory.  
The explicit forms of two-point functions 
give predictions for the large $N$ behavior 
of Matrix theory with some special cutoff. 
We discuss the possible implications of our results for 
the large $N$ dynamics of D-particles and for the Matrix-theory conjecture.  
We find an anomalous scaling behavior 
with respect to the large $N$ limit 
associated to the infinite momentum limit 
in 11 dimensions, suggesting the existence 
of a screening mechanism for the transverse extension 
of the system.

\newpage
\section{Introduction}
The so-called AdS-CFT correspondence 
\cite{maldacena}\cite{gpk}\cite{witten}  
is originated 
from the comparison of two different but dual 
descriptions of 
D-branes, namely as classical BPS solutions of 
effective supergravity in the low-energy limit of closed 
superstring theory and a direct treatment of 
Dirichlet branes as collective degrees of freedom in 
open superstring theory, whose low-energy limit are effectively 
some conformal field theories describing
 the lowest modes of open strings.   
     From the viewpoint of closed string theories or supergravity, 
the system is a field theory in the bulk space-time, while 
in the viewpoint of the world-volume theory of 
Dirichlet branes the system is regarded as a field theory 
defined on a 
boundary of the bulk space-time.   
 
In the most typical case of AdS$_5\times$ S$^5$,  the correspondence has been used to make various 
predictions  for the 
behavior of 3+1 dimensional super conformal Yang-Mills 
theory in the large $N$ limit in its strong 
coupling regime $g_{{\rm YM}}^2N\gg 1$.  
 For example, the masses of field excitations 
around the conformal 
invariant AdS background are related \cite{gpk}
\cite{witten} to the conformal dimensions of the physical operators of super Yang-Mills theory. Indeed the agreement between them is regarded as a strongest piece of 
evidence for the validity 
of the AdS-CFT correspondence.  

It should be stressed that 
 the conformal symmetry plays a crucial role 
 for establishing the 
connection between the correlation functions based on the 
ansatz proposed in \cite{gpk} and \cite{witten}, where 
Yang-Mills theory is assumed to live on the 
boundary of the AdS space-time.  In the original 
discussion in \cite{maldacena}, use of 
Yang-Mills theory is justified as the description 
of D-brane interactions whose distance scales are 
much smaller than the string scale $\ell_s$, 
by fixing the energy $U=r/\ell_s^2$ of open strings 
in the limit of small $\ell_s$, where, 
as is now well known,  the higher string modes 
can be neglected. On the side of 
supergravity, this allows us to take the 
`near-horizon' limit which approximates  
$1+g_sN\ell_s^4/r^4=1+ g_sN/\ell_s^4U^4 $ 
by $g_sN/\ell_s^4U^4$.  However, in the 
prescriptions of \cite{gpk} and \cite{witten}, 
we have to consider the `boundary' of the 
near horizon region by taking the limit $U\rightarrow \infty$. 
Obviously, this tacitly assumes that we can extend 
the correspondence of both descriptions up to the 
region where the near-horizon approximation begins 
to lose its justification,  namely 
$g_sN/\ell_s^4U^4\sim 1$ or 
$r \sim (g_sN)^{1/4}\ell_s \gg \ell_s$. 
This length scale exceeds the naive region of 
validity of the Yang-Mills description for the dynamics of 
open strings.  A possible support for 
this extension of the region seems to be the conformal 
symmetry: Once the correspondence is established at 
sufficiently small distance scales, one can enlarge it to 
larger distance scales  to the extent that  the conformal symmetry is valid on both sides.  The condition 
$r < (g_sN)^{1/4}\ell_s$ places the limit for the validity of 
this assumption from the side 
of supergravity.  Thus the basic assumption 
behind the AdS-CFT correspondence is 
that the correspondence between classical supergravities and 
Yang-Mills matrix models (or any appropriate conformal 
field theories)  is extended to 
the whole `conformal region' 
characterized by the near-horizon condition.  
 Assuming this feature, 
the limit $\ell_s \rightarrow 0$ is not essential and 
we can adopt any convenient unit of length for 
discussing the AdS-CFT correspondence.

The conformal symmetry 
can also be used to constrain \cite{maldacena} the dynamics of (probe) D-branes 
themselves in the AdS background. It has been shown that the 
conformal symmetries on supergravity side and on super Yang-Mills side can be directly related 
 from this point of view. 
For example, the characteristic scale 
$(g_sN)^{1/4}\ell_s$ is shown to be 
obtained \cite{jky1} from the side of Yang-Mills theory 
without relying upon the correspondence with 
supergravity. The deeper meanings 
behind these conformal symmetries have been 
discussed from different viewpoints 
\cite{suss-witten}\cite{li-yo}.  
For instance, 
the space-time uncertainty relation \cite{yo} 
has been the motivation for 
extending the conformal symmetry to general D-branes 
in ref. \cite{jy}.  

It has been discussed \cite{itzak} 
that the  correspondence between 
supergravity and super Yang-Mills theories 
  may be extended to other Dirichlet branes of different dimensions which are not 
necessarily described by conformally  invariant field theory. 
From the viewpoint of the conformal symmetry, it was  argued in \cite{jy} \cite{jky2} 
that a certain extended version of conformal symmetry 
exists for general D-branes. The 
extended symmetry indeed is shown to be 
as effective for constraining 
the dynamics of probe D-branes in the background of 
heavy D-brane sources as in the case of ordinary conformally 
invariant theory.  Some aspects of the 
generalized conformal symmetry 
 have been studied \cite{ref-gen} from a variety of different 
viewpoints.  

The aim of the present work is 
first to substantiate the previous  
discussions of the generalized conformal symmetry 
by establishing the correspondence between the 
excitation spectrum in supergravity around the 
background of a heavy  D0 source 
and the  physical operators in 
supersymmetric quantum mechanical model 
describing D-particles, namely, Matrix theory. 
We confirm that they behave  as 
expected from the generalized conformal 
symmetry.  Secondly, and more importantly, 
supergravity guided by the generalized conformal 
symmetry enables us to predict the 
explicit forms for the correlators of Matrix theory 
operators in a certain special regime of the large 
$N$ limit.  We will discuss  the implications of our 
results for the dynamics of many D-particle systems and 
for  Matrix theory from both the viewpoints of 
discrete light-cone quantization for finite and fixed $N$ 
and of the large $N$ 
infinite momentum frame. We find some 
unexpected anomalous behaviors in the scaling 
properties in the large $N$ limit.  

It might be worthwhile to add a remark here. 
One of the unsolved problems in Matrix theory is 
to establish whether the model is consistent with 
11 dimensional supergravity which is regarded as 
the low-energy description of M-theory.  Since 10 dimensional 
IIA supergravity is the dimensional reduction of 11 dimensional supergravity, one might think that 
invoking supergravity-matrix model correspondence 
 almost amounts to assuming the 
result.  But that is not correct.  
 The Yang-Mils matrix model is intrinsically 
defined only in 10 dimensional space-time, and 
therefore it is not at all evident what is the 
appropriate interpretation of the theory 
from the viewpoint of 11 dimensional M-theory. 
We can in principle test the idea of Matrix theory
 by checking whether or not the behavior 
 of the matrix model in the large $N$ limit predicted by the generalized AdS-CFT correspondence  supports the 
interpretation of the model as being 
defined at the infinite momentum frame in 11 dimensional space-time.   
 
The paper is organized as follows. 
In section 2, we briefly summarize the M-theory 
interpretation of the D0-brane matrix model  and 
review the 
generalized conformal symmetryz: Although most of the 
points there 
 have  already been 
discussed in previous works \cite{jy, jky1, jky2},  
we hope that our discussion is 
useful for the purpose of clarification.  We then 
analyze the possible regime of validity 
in which we can expect the correspondence between 
supergravity and the Yang-Mills description of D-particles 
in the large $N$ limit.  In particular, we 
emphasize that we have to set 
an infrared cutoff of the order $(g_sN)^{1/7}\ell_s$ 
in the strong coupling region $g_sN> 1$ of the large $N$ limit 
in applying the correspondence. 
 In section 3, we present the results of a complete harmonic 
analysis of the bosonic excitations around the classical 
D-particle solution in type IIA supergravity in 10 dimensions. 
In section 4, using the results obtained in the 
previous section, we discuss the correspondence of the 
matrix model operators with the supergravity fluctuations, 
and the possible implications 
of the results from the viewpoint of Matrix-theory 
interpretation.  In Appendix, we give a brief summary 
of the definitions and basic properties of the general 
tensor harmonics. 

\section{Generalized conformal symmetry and the  large $N$ limit in D0-matrix model}
Let us start from the standard matrix model action for 
D-particles
\EQ
S =\int dt \, \Tr 
\Bigl( {1\over 2g_s\ell_s} D_t X_i D_t X_i + i \theta D_t \theta 
+{1 \over 4g_s\ell_s^5} [X_i, X_j]^2 -
{1\over \ell_s^2}\theta \Gamma_i [\theta, X_i]\Bigr) 
\label{action}
\EN
where $X_i \, \, (i=1,2,\ldots, 9)$ are the $N\times N$ hermitian matrices 
representing the collective modes of $N$ D-particles 
coupled with the lowest modes of open strings.  
In the present paper,  we will use the $A=0$ gauge. 
The diagonal elements of the coordinate matrices $X_i$ 
are interpreted as the 9 dimensional spatial positions of the 
D0-branes in 10 dimensional space-time. 
Alternatively, the same action can be interpreted 
\cite{bfss} as the 
effective action for the lowest KK-mode of the graviton 
supermultiplet in 11 dimensional space-time 
in the infinite-momentum frame where the light-like 
momentum $P_-={1\over 2}(P_0-P_{10})$ is quantized by the unit $1/R$ with 
$R=g_s\ell_s$ being the compactification radius 
along the 11th (space-like) dimension.    For the 11 dimensional 
interpretation, it is more natural to rewrite the Lagrangian as 
\EQ
L = \Tr 
\Bigl( {1\over 2R} D_t X_i D_t X_i + i \theta^T D_t \theta 
+{R \over 4\ell_P^6 } [X_i, X_j]^2 -
{R\over \ell_P^3}\theta^T \gamma_i [\theta, X_i]\bigr) ,
\EN
and also the corresponding Hamiltonian as  
\EQ
H\, (=-2P^-) =R\Tr h={N\over P_-}\Tr \, h
\label{Hamilton}
\EN
\EQ
h={1\over 2}\Pi^2  - {1\over 4\ell_P^6}[X^i, X^j]^2 
+ {1\over 2\ell_P^3}
[\theta_{\alpha},[X^k, \theta_{\beta}]]\gamma^k _{\alpha\beta}
\label{hamiltonian}
\EN
using the 11 dimensional Planck length $\ell_P=g_s^{1/3}\ell_s$.  For any fixed finite $g_s$, 
the infinite momentum limit $P_-\rightarrow 
\infty$ requires to 
take the large $N$ limit. The form of the Hamiltonian 
implies that for the infinite momentum limit 
for finite $g_s$ to be meaningful, the 
spectrum of the operator $\Tr \, h$ in the large $N$ limit 
must scale as $1/N$ in the low-energy (near threshold) region. 
In other words, the time must scale as $N$. Therefore 
it is important to analyze the large-time behavior of 
appropriate correlators for the study of the Matrix-theory 
conjecture.  
The usual discussion \cite{seiberg} for justifying Matrix theory 
at finite $N$ \cite{suss}  only deals with the (spatial) 
length scale smaller than
 the string scale.  For example, such an argument is 
insufficient to explain the results \cite{bbpt} \cite{okawayoneya} of 
the explicit computations of D-particle scattering beyond 
one-loop approximation, in which the perturbative approximation becomes better and better 
 for larger distance scales. 
One of the most crucial issues of Matrix theory is 
to understand whether the theory can 
describe gravity consistently at large distance 
scales  either in the finite $N$ or in the large $N$ limit.  

The space-time scaling 
property of D-particles \cite{dkps} can be qualitatively 
summarized into an uncertainty relation in space-time 
\cite{yo} 
\EQ
\Delta T \Delta X \sim \ell_s^2 , 
\label{stur}
\EN
between the minimum 
uncertainties $\Delta T$ and $\Delta X$ 
with respect to time and space, respectively. 
This relation is invariant under the opposite scaling 
of time and space
\EQ
X_i(t) \rightarrow X_i'(t') =\lambda X_i(t) , \quad t\rightarrow t'=\lambda^{-1}t \, .
\label{oppscaling}
\EN
The matrix model action is invariant under these scaling 
transformations if the string coupling is scaled as 
\EQ
g_s \rightarrow g_s'=\lambda^3 g_s .
\label{couplingscaling}
\EN
The scaling of the string coupling is just equivalent to 
the fact that the characteristic spatial and time scales of the 
theory are $g_s^{1/3}\ell_s=\ell_P $ and $g_s^{-1/3}\ell_s$, 
respectively,  
apart from dimensionless but string-coupling 
independent proportional constant. 
This  can be derived by combining the space-time uncertainty relation with the ordinary quantum mechanical uncertainty relations.  The space-time uncertainty relation 
may be regarded as a simple but universally valid 
 underlying principle \cite{yo} 
governing the short distance space-time 
structure of string/M theory.  
In \cite{jy}, we have pointed out that the symmetry 
of the model and the space-time uncertainty relation  can be 
extended to the special conformal transformation,  
\EQ
\delta_K X_i = 2  t X_i ,   
 \delta_K t =-  t^2 , \, \, \delta_K g_s =6  t g_s . 
\label{eq29}
\EN
These symmetries are appropriate to be called as 
`generalized conformal symmetry'. 

The same scaling properties exist on the side of 
supergravity solution too. From the point of view of 10 dimensional type IIA theory, the solution is expressed as 
\EQ
ds_{10}^2 = -e^{-2\tilde{\phi}/3}dt^2 + e^{2\tilde{\phi}/3}dx_i^2 .
\label{d0metric}
\EN
\EQ
e^{\phi}= g_s e^{\tilde{\phi}}
\label{dilaton}
\EN
\EQ
e^{\tilde{\phi}} =\bigl( 1 +{q\over r^7}\bigr)^{3/4}
\EN
\EQ
A_0=-{1\over g_s}\bigl( 
{1\over 1+{q\over r^7}}-1\bigr)
\bigr) , 
\EN
and the charge $q$ is given by 
$
q=60\pi^3 (\alpha')^{7/2} g_sN
$
for $N$ coincident D-particles.  
In the near horizon limit $q/r^7\gg 1$ where 
the factor $1+q/r^7$ is replaced by $q/r^7$,  
the metric, dilaton and the 2-form field strength $F_{i0}dx^i\wedge dt\propto 7r^6 dr\wedge dt/g_s^2$ are all invariant under the scale and the special conformal transformations
\EQ
r\rightarrow \lambda r, \, 
t \rightarrow \lambda^{-1}t ,\, 
g_s \rightarrow \lambda^3 g_s ,
\label{sugrascale}
\EN
 \EQ
\delta_K t = - \epsilon (t^2 +{2q\over 5 r^5}) , 
 \quad \delta_K r =2 \epsilon tr ,
\quad \delta_K g_s=6 \epsilon t g_s 
\label{sugraspecial}
\EN
which together with time translation form an 
$SO(1,2)$ algebra.  
The additional term in the special conformal transformation 
does not affect the space-time uncertainty relation, 
but plays an important role  \cite{jy} in constraining the 
effective action of a probe D-particle in the 
background metric (\ref{d0metric}). 
The mechanism how the additional term 
${2q\over 5 r^5}$ emerges in the bulk theory 
was clarified in refs. \cite{jky1,jky2} for general case 
of D$p$ -branes from the point of view of matrix models, 
namely, from the boundary theory.  It should be emphasized 
that  the new 
scale $q^{1/7}\propto (g_sN)^{1/7}\ell_s $ which characterizes the radius of the 
near horizon region around the system of the source D-particles 
was thus derived entirely within the logic of 
Yang-Mills matrix models.  This may be regarded as 
independent evidence for the dual correspondence between 
supergravity and Yang-Mills matrix models.

As noted in previous works,  the combination 
$g_s/r^3$ is invariant under the generalized conformal transformation, 
and hence the D0 space-time metric 
can be regarded as a sort of `quasi' AdS$_2
\times$ S$^8$ with a {\it variable but conformally 
invariant} radius  proportional to $\rho
=(g_sN\ell_s^3/r^3)^{1/4}\ell_s
\propto (q\e^{\tilde{\phi}})^{1/7}$. 
Equivalently, the D0 metric in the near 
horizon limit is related 
by  a Weyl transformation  to the true AdS$_2\times$ S$^8$ as 
\EQ
ds_{10}^2
=(q\e^{\tilde{\phi}})^{2/7}\Bigl[ -\bigl({2\over 5}\bigr)^2 {dt^2- dz^2\over z^2}
+d\Omega^2_8\Bigr]\equiv \e^{2\tilde{\phi}/7}
ds_{AdS}^2 ,
\label{weyltransformation} 
\EN
\EQ 
z\equiv 2q^{1/2}r^{-5/2}/5.
\label{rtoz}
\EN 
 This representation will 
be technically useful for performing the 
harmonic analysis in the next section.  
However, we have to be careful in considering the 
generalized conformal transformation 
using this metric, since the coordinate transformation 
(\ref{rtoz}) involves $q$ which is not constant 
under the conformal transformation. 
In particular, the symmetry is not the 
isometry under any coordinate transformation.  
For example, taking derivative and performing 
the generalized conformal transformation 
are not commutative. 
To avoid possible confusions caused by this subtlety, it is 
safe to return to the original coordinate $r$ whenever 
we discuss the conformal transformation of the 
fields. 
Note that in this picture the Weyl noninvariance of the 
theory, or the nonconformal 
nature of the theory,  is canceled by the transformation of the string coupling constant. In other words, the 
Weyl factor itself is treated as  being invariant under the 
generalized conformal transformation.  
 
Whether the transformations being accompanied by 
the change of the 
string coupling constant should be called as a symmetry transformation may perhaps be a matter of debate.    
Our viewpoint is that the above transformations 
must be interpreted in the whole moduli space of the 
vacua of perturbative string/M  theory. The change of the 
string coupling (namely,  the asymptotic value of the dilaton 
expectation value or the asymptotic 
value of the compactification radius along the 
11th dimension) is 
interpreted as a change of the vacuum at infinity. 
The conformal transformation is a symmetry characterizing the short distance property of the theory, but it must 
be accompanied by a change of vacuum at large 
distance asymptotic region, depending on 
the D-brane states we are considering. 
Note that the linear time dependence of the dilaton 
to its first order is an allowed deformation 
of the vacuum in the flat space-time. 
Ultimately, the string coupling should be 
eliminated from the fundamental 
nonperturbative formulation of the 
theory. 

It is also worthwhile to point out that the scaling transformation of the generalized conformal 
symmetry is related to the discrete light-cone 
interpretation of the model with fixed $N$ in the sense that 
the transformation is equivalent with the boost 
transformation, $t\rightarrow \lambda^{-2} t, 
R\rightarrow \lambda^2 R, X_i \rightarrow X_i, \ell_P \rightarrow \ell_P$ and $\ell_s \rightarrow \lambda^{-1}\ell_s$,  
when we use the unit such that the 
11 dimensional Planck length is fixed by changing the 
unit of length globally.  
Alternatively, we can use the unit such that the 
time is not changed. This leads to 
the transformation 
$t\rightarrow t, R\rightarrow \lambda^4 R, X_i \rightarrow \lambda^2 X_i, \ell_P 
\rightarrow \lambda^2\ell_P$ and $\ell_s \rightarrow \lambda \ell_s$. 
The latter transformation is nothing but the 
redefinition proposed in \cite{seiberg}. 

Given now 
that there exists the same generalized 
conformal symmetry on 
both sides of the matrix model and supergravity, 
we can define the conformal dimensions of 
operators on both sides and consider the 
correspondence between the spectra of both 
sides by following the familiar 
prescription \cite{gpk} \cite{witten} 
 of computing  the correlators of the matrix 
model using supergravity. 
Let us analyze the region of validity 
for the correspondence.  

Throughout the present paper, we will use the unit of length in which the string 
scale $\ell_s$ is {\it fixed}, instead of the 
familiar convention of fixing the 
energy of the open strings stretched among 
D-branes.  
First  for the supergravity approximation to be good, the curvature radius $\rho$ must be 
larger than the string scale giving 
\EQ
r \ll (g_sN)^{1/3}\ell_s, 
\label{smallcurvature}
\EN
while the near horizon condition gives 
\EQ
r\ll (g_sN)^{1/7}\ell_s \, 
\label{nearhorizon}
\EN
If we further require that the effective  string coupling which can be determined  locally by the 
dilaton is small, we must have also 
\EQ
g_s^{1/3}N^{1/7}\ell_s =g_s^{4/21}(g_sN)^{1/7} \ell_s\ll r
\label{smallcoupling} 
\EN
which essentially demands that the effective local compactification radius along 11th dimensions must be smaller than the string scale. This is necessary as long as we use the 
10 dimensional supergravity picture.   Since the last two conditions (\ref{nearhorizon})  and 
(\ref{smallcoupling}) both have the same behavior 
with respect to $N$, the existence of finite 
region of validity for the correspondence always require that the string coupling constant must be very small $g_s\ll 1$.    
This is especially true if one wishes to justify the 
Yang-Mills description of open strings 
which requires to go to much shorter distances than 
the string scale.  Note that the characteristic length 
scale of elementary processes of D-particle interactions 
is always expected to be $g_s^{1/3}\ell_s$. 
On the other hand, the first two conditions 
tell us that the extent to which the region of 
validity extends in the large distances depends 
crucially on whether 
$
g_sN >1
$ or
$
g_sN <1
$.
In the former strong coupling region, we have to put the 
infrared cutoff at $r \sim (g_sN)^{1/7}\ell_s$, while in the 
the latter weak coupling region that is 
$r\sim (g_sN)^{1/3}\ell_s$. 
  From the point of view of  Matrix theory in which we 
wish to investigate the large-$N$ dynamics for fixed 
$g_s$, 
the former restriction is problematical, since it is expected from a general argument \cite{pol}  based on the virial theorem that 
the typical extension of the system of $N$ D-particles 
is $(g_sN)^{1/3}\ell_s (\gg (g_sN)^{1/7}\ell_s$) for large $g_sN$. 
We will come to this question later. 
To summarize, 
we must be in the region of 
small string coupling constant and large $N$ 
such that $g_sN\gg 1$, in order to go 
beyond the string scale $r> \ell_s$ on the basis 
of the AdS-CFT type correspondence between supergravity 
and Yang-Mills matrix models.  It is also important to recall the remark which was already stressed in 
the Introduction, that 
we consider  the whole near-horizon region 
(or conformal region) till its limit 
characterized by $(g_sN)^{1/7}\ell_s\gg \ell_s$.   
This is also necessary 
for studying the Matrix-theory conjecture. 
On the other hand, 
as far as we remain in the weak coupling region 
$g_sN < 1$, 
the region of validity of the supergravity-matrix model 
correspondence is restricted in the region below the string scale even if we take the large $N$ limit.  

 If we take the viewpoint of 11 dimensional M-theory and fix the Planck length 
$\ell_P=g_s^{1/3}\ell_s$, the near horizon condition 
is $r \ll g_s^{-4/21}N^{1/7}\ell_P$ which contains the 
small curvature region $r\ll N^{1/3}\ell_P$ for sufficiently small string coupling 
for any {\it fixed} $N$. However, in the large $N$ limit, 
these two infrared conditions can compete 
depending on the magnitude of $g_sN$ as explained 
above. 
 Remember that in the 
case of D3 brane the infrared cutoff 
associated with the small curvature condition 
and that coming from  the near horizon condition  are governed by the 
same scale $r\sim (g_sN)^{1/4}
\ell_s \gg \ell_s$, in sharp contrast to the case of D-particles.  
In passing, 
the condition that the  nonvanishing components of 11 dimensional Riemann tensor ($\sim 
\partial_i\partial_j (q/r^7)$) 
around the D0-metric  are small compared to
 the characteristic curvature $\ell_P^{-2}$  is
\EQ
 g_s^{5/27}N^{1/9}\ell_s \ll r
\label{11dcurvature}
\EN
which is weaker than 
the 10 dimensional condition (\ref{smallcoupling}) 
only when $N$ is bigger than $\sim g_s^{-14/3}$. 
Therefore we have to keep in mind that unless the 
latter  condition is satisfied it could be dangerous to 
elevate the classical 10 dimensional supergravity to 
classical 11 dimensional supergravity. 
In particular, this shows that in the ' t Hooft limit in which we fix $g_sN$ in taking the limit $N\rightarrow \infty$, 
we cannot lift up the theory to 11 dimensions, 
while, if we take the large $N$ limit with a fixed but small $g_s$, the 11 dimensional picture enables us to go beyond the 
10 dimensional lower bound (\ref{smallcoupling}).  
This will be relevant for the discussion of the 
Matrix-theory conjecture at the end of the present paper.

 \section{Harmonic analysis on D-particle background in type IIA supergravity}
\setcounter{equation}{0}
The idea behind the
prescription \cite{gpk} \cite{witten} for computing the correlators
of the Yang-Mills theory using supergravity
is as follows. Suppose we study the
system of a heavy source of $N$ D-branes
by putting a light D-brane far away ({\it i.e., }
outside the near horizon region) from the
source as a probe.  Thus we are considering the
U($N+1$) Yang-Mills theory broken down to
U($N$)$\times$U(1).
On the side of supergravity,  the effect of the probe
can be treated as a perturbation around the
background of the source D-branes. For the inner region
near to the horizon,
information on a given state of the probe
can be encoded into the boundary condition
$\{h_i\} $ for the independent set of the
perturbing fields  at the boundary of the
near horizon region.  From the Yang-Mills viewpoint,
on the other hand, the perturbing fields produced by the
probe should be coupled at the boundary  to
some independent set of
 operators of the Yang-Mills theory describing the
source D-branes.  This leads to the ansatz for the
correlators of the Yang-Mills operators in Euclidean metric, as
\EQ
\e^{-S_{SG}[h]} = \langle \exp \{\int dt \sum_i h_i (t){\cal O}_i(t)\}
\rangle_{U(N)} ,
\label{cprescription}
\EN
where $S_{SG}[h]$ is the supergravity action
as the functional of the boundary value of the
perturbing fields. Although it is not literally
correct to
say that either the source or probe D-branes do
reside  at the boundary, we can interpret
the relation (\ref{cprescription}) {\it as if} 
the U($N$) Yang-Mills theory  describing the source
D-branes lives on the boundary.
The conformal symmetries on both sides
allow us to diagonalize the operators and the
perturbing fields with respect to the
conformal dimensions.

In this section, we perform the
diagonalization of the perturbing fields on the side of
supergravity by carrying out the harmonic analysis
of the linearized perturbations around the D0 background.
The calculation follows essentially the same method as in the case of 
non-dilatonic branes.
See \cite{krn} for the typical case of $AdS_5\times S^5$.
In the present paper, we only treat the bosonic 
perturbations.  The reader who cannot be patient  in following  
the details of the harmonic analysis might want to skip 
the rest of 
this section and directly go to section 4 where we discuss 
the correspondence between the spectrum of the 
fluctuations and the Matrix model operators and 
its implications. 

Let us start from the standard IIA supergravity action in 10
dimensions
using the string frame ($\kappa^2 \sim g_s^2 \ell_s^8$),
\EQ
S=
{1\over 2\kappa^2}
\int dx^{10}\sqrt{-g}
\Bigl[
\e^{-2\tilde{\phi}}\Bigl(
R + 4(\partial \phi)^2 -{1\over 2}|H_3|^2
\Bigr)
-{g^2_s\over 2}|F_2|^2 -{g^2_s\over 2}|\tilde{F}_4|^2
\Bigr]-
{g^2_s\over 4\kappa^2}\int B_2\wedge
F_4\wedge F_4
\EN
where we have written down only the
bosonic part of the action and
$H_3=dB_2, F_4=dA_3, \tilde{F}_4=F_4 -A_1 \wedge H_3,
F_2=dA_1$.  Note also that $|F_p| ^2
=F_{\mu_1\mu_2\cdots \mu_p}
F^{\mu_1\mu_2\cdots \mu_p}/p!$. The nontrivial background fields
exist for the metric, dilaton and RR 1-form fields.
Although this is not compulsory,
\footnote{We have explicitly
confirmed that the final results for the
dimensions and correlators are
the same as those obtained using the
metric without making the Weyl
transformation. } it is convenient
for the harmonic analysis to
make the Weyl transformation (\ref{weyltransformation})
for the metric. We denote the metric of the transformed
frame by $g^{(total)}_{\mu\nu}$ and decompose the fields as
$g^{(total)}_{\mu\nu}=g_{\mu\nu}+ h_{\mu\nu}$.
Namely, we denote the background metric in the AdS frame
by $g_{\mu\nu}$ and the fluctuation by $h_{\mu\nu}$.
For other fields similarly, we denote the
background fields by using the original notation
as $\tilde{\phi}, A_{\mu}$ and the fluctuations
around them by putting `hat'  such as $\hat{\phi}, \hat{A}_{\mu}$.
After the Weyl transformation the action takes the form
\EQA
S&=&{1\over 2\kappa^2}\int d^{10}x\sqrt{-g}\left[ \e^{-{6
\over 7}
\tilde{\phi}}\left\{ R+{16\over 49}\partial_\mu\phi\partial^\mu\phi\right\}
-{1\over 12}\e^{-{10\over 7}\tilde{\phi}}H_{\mu\nu\rho}H^{\mu\nu\rho}
\right.\nonumber\\
&& \hspace{-1.5cm}-{g^2_s\over 4}\e^{{6\over 7}\tilde{\phi}}F_{\mu\nu}F^{\mu\nu}  -{g^2_s\over
48}\e^{{2\over 7}\tilde{\phi}}F_{\mu\nu\rho\sigma}F^{\mu\nu\rho\sigma}
\left. -{g^2_s\over 2\cdot 2!4!4!}\epsilon^{\mu_1\cdots\mu_{10}}B_{\mu_1\mu_2}
F_{\mu_3\mu_4\mu_5\mu_6}F_{\mu_7\mu_8\mu_9\mu_{10}}\right].
\EQN

\subsection{The fluctuations of  metric, dilaton and RR 1-form fields}
We first consider the fluctuations of the metric, dilaton and
RR 1-form field, since other fields without the nontrivial
background fields are decoupled from them.
The linearized equations for the fluctuations are
listed below.
\EQA
&&\hspace{0.6cm} {{{h^\mu}_\mu}^{;\nu}}_{;\nu}  
-{{{h^\mu}_\nu}^{;\nu}}_{;\mu}+{8\over 21}(2 {{h^{\nu}}_\mu}^{;\mu}
-{{h^\mu}_\mu}^{;\nu})\partial_\nu{\tilde{\phi}}\nonumber\\
&&\hspace{-2.1cm}+{h^{\mu}}_\nu\left\{ {R^\nu}_\mu+{16\over  
21}D_\mu\partial^\nu{\tilde{\phi}} -{48\over  
147}\partial^{\nu}\tilde{\phi}\partial_\mu{\tilde{\phi}}
+{1\over 2} g^2_s \e^{{12\over 7}\tilde{\phi}}  
F_{\mu\rho}F^{\nu\rho}\right\}-{16\over 21}D_{\mu}\partial^{\mu}\hat{\phi}
+{96\over 147}\partial_\mu{\tilde{\phi}}\partial^\mu\hat{\phi}\nonumber\\
&&\hspace{-1.6cm}+{6\over 7}\hat{\phi}\left\{ R+{16\over  
21}D^\mu\partial_\mu{\tilde{\phi}} -
{48\over 147}\partial_\mu{\tilde{\phi}}\partial^\mu{\tilde{\phi}}
-{1\over 2} g^2_s\e^{{12\over 7}\tilde{\phi}}  
F_{\mu\nu}F^{\mu\nu}\right\}-{1\over 2} g^2_s\e^{{12\over 7}\tilde{\phi}}  
\hat{F}_{\mu\nu}F^{\mu\nu}=0
\label{d}
\EQN
\EQA
&&\hspace{-2cm} -{1\over 2}({{{h^\rho}^\mu}^{;\nu}}_{;\rho}
+{{{h^\rho}^\nu}^{;\mu}}_{;\rho}
-{{{h^\mu}^\nu}^{;\rho}}_{;\rho}
-{{{h^\rho}_\rho}^{;\mu}}^{\nu})+{1\over 2} g^{\mu\nu}(  
{{{h^\rho}_\sigma}^{;\sigma}}_{;\rho}-
{{{h^\rho}_\rho}^{;\sigma}}_{;\sigma})+{3\over  
7}({{{h^\sigma}^\mu}^{;\nu}}+{{{h^\sigma}^\nu}^{;\mu}}
-{{{h^\mu}^\nu}^{;\sigma}})\partial_\sigma{\tilde{\phi}}\nonumber\\
&&\hspace{-0.8cm}-{3\over7}g^{\mu\nu}(2{{{h^\sigma}_\rho}^{;\rho}}
-{{{h^\rho}_\rho}^{;\sigma}})
\partial_\sigma{\tilde{\phi}} +h^{\mu\nu}\left\{ -{1\over 2} R -{6\over  
7}D^\rho\partial_\rho{\tilde{\phi}} +{4\over  
7}\partial^\rho{\tilde{\phi}}\partial_\rho
{\tilde{\phi}} +{1\over 8}g^2_s\e^{{12\over 7}\tilde{\phi}}  
F_{\rho\sigma}F^{\rho\sigma}\right\}\nonumber\\
&&\hspace{0.5cm}+{h^\mu}^{\rho}\left\{
{R_\rho}^\nu +{6\over 7}D_\rho\partial^\nu{\tilde{\phi}}
-{20\over 49}\partial^\nu{\tilde{\phi}}\partial_\rho{\tilde{\phi}}
-{1\over 2} g^2_s\e^{{12\over 7}\tilde{\phi}}  
F_{\rho\sigma}F^{\nu\sigma}\right\}\nonumber\\
&&\hspace{0.5cm}+{h^\nu}^{\rho}\left\{
{R_\rho}^\mu +{6\over 7}D_\rho\partial^\mu{\tilde{\phi}}
-{20\over 49}\partial^\mu{\tilde{\phi}}\partial_\rho{\tilde{\phi}}
-{1\over 2} g^2_s\e^{{12\over 7}\tilde{\phi}}  
F_{\rho\sigma}F^{\mu\sigma}\right\}\nonumber\\
&&\hspace{-1cm}-{1\over 2} g^2_s\e^{{12\over  
7}\tilde{\phi}}{h^\rho}^{\sigma}{F^\mu}_\rho
{F^\nu}_\sigma+g^{\mu\nu} {h^\rho}_\sigma\left\{ -{1\over 2} {R^\sigma}_\rho  
-{6\over 7}D^\sigma\partial_\rho{\tilde{\phi}} +{4\over  
7}\partial^\sigma{\tilde{\phi}}\partial_\rho
{\tilde{\phi}} +{1\over 4}g^2_s\e^{{12\over 7}\tilde{\phi}}  
F_{\rho\eta}F^{\sigma\eta}\right\}\nonumber\\
&& \hspace{-0.4cm}-{6\over 7}D^\mu\partial^\nu\hat{\phi}
+{6\over 7}g^{\mu\nu} D^\rho\partial_{\rho}\hat{\phi}+{20\over  
49}\partial^\mu\hat{\phi}\partial^\nu{\tilde{\phi}}
+{20\over 49}\partial^\mu{\tilde{\phi}}\partial^\nu\hat{\phi}
-{8\over 7}g^{\mu\nu}\partial^\rho\tilde{\phi}\partial_\rho\hat{\phi}\nonumber\\
&&\hspace{0.6cm} +{6\over 7}\hat{\phi} (R^{\mu\nu} +{6\over  
7}D^\mu\partial^\nu{\tilde{\phi}}
-{20\over 49}\partial^\mu{\tilde{\phi}}\partial^\nu{\tilde{\phi}} +{1\over  
2} g^2_s\e^{{12\over 7}\tilde{\phi}}
{F^\mu}_\rho F^{\nu\rho}) \nonumber\\
&&\hspace{0.6cm} +{6\over 7}\hat{\phi}g^{\mu\nu} (-{1\over 2} R-{6\over  
7}D^\rho\partial_\rho{\tilde{\phi}}
+{4\over 7}\partial_\rho{\tilde{\phi}}\partial^\rho{\tilde{\phi}} -{1\over  
8} g^2_s\e^{{12\over 7}\tilde{\phi}}
F_{\sigma\rho} F^{\sigma\rho})\nonumber \\
&&\hspace{1cm} +g^2_s \e^{{12\over 7}\tilde{\phi}} \left\{ {1\over  
2}{\hat{F}^\mu}_{\;\;\rho}
F^{\nu\rho}+{1\over 2}{\hat{F}^\nu}_{\;\;\rho}
F^{\mu\rho}-{1\over 4}g^{\mu\nu} \hat{F}_{\rho\sigma}
F^{\rho\sigma}\right\}=0
\label{h}
\EQN
\EQA
&&\hspace{-1.8cm}-{{h^\rho}_\sigma}^{;\sigma} {F^\mu}_\rho
+{1\over 2}{{h^\sigma}_\sigma}^{;\rho} {F^\mu}_\rho
-h^{\mu\rho ;\sigma}F_{\rho\sigma}+{h^{\rho}}_\sigma\left\{ -D_\rho  
F^{\mu\sigma}-{6\over 7}
\partial_\rho{\tilde{\phi}} F^{\mu\sigma}\right\}\nonumber\\
&&+{6\over 7}\partial_\rho\hat{\phi} F^{\mu\rho}+D_\rho {\hat{F}}^{\mu\rho}  
+{6\over 7}\partial_\rho{\tilde{\phi}}
{\hat{F}}^{\mu\rho} =0
\label{a}
\EQN
They are obtained by the variations
of dilaton, metric, and RR 1-form fields, respectively.

   From now on unless otherwise specified, we use the following conventions for denoting the tensor  
indices:  $\mu, \nu, \ldots $ for full 10 dimensional
contractions. $i, j, \ldots$ for the metric on the
sphere S$^8$. Thus, for example, the covariant derivative
$D_i$  is defined using only the
metric of S$^8$.  The convenience of the AdS frame metric
$ds^2_{AdS}$ is that the S$^8$ metric is
completely decoupled from that of AdS$_2$.
Note however that our background is
{\it never} the AdS$_2 \times$ S$^8$ itself:
the Weyl factor induces complicated couplings
between various modes, in contrast to the case of
`non-dilatonic' D-branes where we can link the 
computations to the representation theory of 
(super) conformal algebra, and makes our analysis
fairly nontrivial. Unfortunately, no analysis has 
been done on the group-theoretic aspect of the 
generalized (super) conformal symmetry. 

To analyze the spectrum it is necessary to fix the
gauge. We adopt the following gauge conditions anticipating the simplicity of the harmonic expansion, 
\EQA
&&D_i({h^i}_j -{1\over 8}\delta^i_j {h^k}_k) =0 ,\nonumber\\
&&D^i h^0_i=D^ih^z_i =0 ,
\label{gaugecond1}\\
&&D^i\hat{A}_i=0 \nonumber .
\EQN
The reader should refer to Appendix for the 
connection of these gauge conditions and the 
harmonic expansion. 
The number of the fields are originally 1(dilaton) +55
(metric)+10(RR 1-form).  There are
10(metric)+1(RR 1-form) constraints coming from the
field equations. The gauge conditions (\ref{gaugecond1})
eliminate $8+2+1=11$ components. Thus we have
$66-2\times 11=44$ physical components.
The result of the harmonic analysis on S$^8$ will show that
for generic value of
angular momentum there is one radial degree of freedom for the symmetric
divergenceless and traceless tensor, two
for the divergenceless vector and three for scalar,
which indeed sum up to 44 total degrees of freedom.

 Using the standard general theory of tensor spherical  
harmonics,
we can expand the fluctuations as
\[
{h^0_0}(x^\mu) =\sum {b^0_0} (t,z) Y(x^i) , \quad 
{h^0_z}(x^\mu) =\sum {b^0_z} (t,z) Y(x^i) ,
\]
\[
{h^z_z}(x^\mu) =\sum {b^z_z} (t,z) Y(x^i) , \quad
{h^i_i}(x^\mu) =\sum b^i_i (t,z) Y(x^i)  ,
\]
\[
\hat{A}_0(x^\mu) =\sum a_0(t,z) Y(x^i) , \quad 
\hat{A}_z(x^\mu) =\sum a_z(t,z) Y(x^i)  ,
\]
\EQ
\hat{\phi}(x^\mu) =\sum \varphi(t,z) Y(x^i)  ,
\label{harms}
\EN

\[
h^0_i(x^\mu) =\sum b^0 (t,z) Y_i(x^i) , \quad 
h^z_i(x^\mu) =\sum b^z (t,z) Y_i(x^i)  ,
\]
\EQ
\label{harmv}\\\hat{A}_i(x^\mu) =\sum a(t,z) Y_i(x^i) ,
\EN
\EQA
&&{h^i}_j(x^\mu) -{1\over 8}\delta^i_j {h^k}_k(x^\mu) =\sum b (t,z)  
Y_{ij}(x^i) .
\label{harmt}
\EQN
We note that the gauge conditions (\ref{gaugecond1}) eliminate the
possible contributions coming from the derivatives of the
harmonics with lower number of tensor indices.
Here, $Y, Y_i$ and $Y_{ij}$ are the scalar, vector and
symmetric-traceless tensor harmonics of $S^8$, respectively.
Note also that we have suppressed the angular momentum index $\ell$ in the harmonic expansion. See Appendix.   

It is appropriate to classify the
coefficient functions in these expansions
into the following three categories :
(1) scalar components  (\ref{harms}),
(2) vector components  (\ref{harmv})
and (3) tensor components (\ref{harmt}).
Since these harmonic functions are mutually independent
(orthogonal to each other),
the linearized equations can be separated into
the coefficient equations for  each harmonic functions.
There arise 16 coefficient equations.  The equation for
tensor components $b$ is a single equation
which is the symmetric traceless tensor part of the
equation (\ref{h})
\EQ
\left[ -{25\over 8}z^2\partial_0\partial_0 b + {25\over  
8}z^2\partial_z\partial_z b -{45\over 8}z\partial_z
 b +({\lambda_T\over 2} -1)b \right]Y_{jk}=0 \, ,\qquad\ell\ge2.
\label{symmtraceless}
\EN
For the vector components $(b^0, b^z, a)$, we have four equations;  three of  
them come
from (\ref{h}) and one from (\ref{a}),   given as
\EQA
&&\hspace{-1cm} {25\over 8}\partial_z\partial_z b^0+{25\over  
8}\partial_z\partial_0 b^z
-{95\over 8}{1\over z}\partial_z b^0 -{45\over 8}
{1\over z}\partial_0 b^z\nonumber\\
&&\quad +\left({\lambda_V\over 2}+21\right){1\over z^2}b^0
-{7\over 2}\left({2\over 5}\right)^{-{19\over 5}}g_s q^{-{2\over  
5}}z^{9\over 5}
\partial_z a =0 ,\qquad\ell\ge 1,
\label{h0iv}
\EQN
\EQA
&&\hspace{-0.5cm}
-{25\over 8}\partial_z\partial_0 b^0 -{25\over 8}\partial_0\partial_0 b^z
+{50\over 8}{1\over z}\partial_0 b^0\nonumber\\
&&\quad +\left( {\lambda_V\over 2}+{7\over 2}\right){1\over z^2} b^z   
+{7\over 2}\left({2\over 5}\right)^{-{19\over 5}}g_s q^{-{2\over  
5}}z^{9\over 5}
\partial_0 a =0 ,\qquad\ell\ge 1,
\label{hziv}
\EQN
\EQ
 -{1\over 2}\partial_0 b^0 -{1\over 2}\partial_z b^z +{19\over  
10}{1\over z}
b^z =0 ,\qquad\ell\ge 2,
\label{hijv}
\EN
\EQA
&&\hspace{-0.5cm}  7{1\over {g_s}}\left({2\over 5}\right)^{9\over  
5}q^{2\over 5}z^{-{19\over 5}}
\left\{ \partial_z b^0 +\partial_0 b^z -{2\over z}b^0\right\}\nonumber\\
&& +{25\over 4}\partial_0\partial_0 a -{25\over 4}\partial_z\partial_z a
-{45\over 4}{1\over z}\partial_z a +(-\lambda_V +7){1\over z^2} a =0,\qquad\ell\ge 1.
\label{aiv}
\EQN
There are remaining 11 coefficient equations for 7
scalar components $(b^0_0, b^0_z, b^z_z, b^i_i,
a_0, a_z, \phi)$; the first of them comes from
the dilaton equation (\ref{d})

\EQA
&& {75\over 14}\partial_z\partial_z b^0_0 -{135\over 14}{1\over z}\partial_z  
b^0_0 -21 {1\over z^2} b^0_0
-{75\over 14}\partial_0\partial_0 b^z_z +{135\over 14}{1\over z}\partial_z  
b^z_z -48 {1\over z^2} b^z_z -{75 \over 7}\partial_0\partial_z b^0_z  
+{135\over 7}{1\over z}\partial_0 b^0_z\nonumber\\
&&-{75\over 14}\partial_0\partial_0 b^i_i +{75\over 14}\partial_z\partial_z  
b^i_i -{30\over 7}{1\over z}\partial_z b^i_i +6 {1\over z^2} b^i_i  
+{200\over 49}\partial_0\partial_0 \varphi -{200\over  
49}\partial_z\partial_z \varphi +{360\over 49}{1\over z}\partial_z \varphi  
+36 {1\over z^2} \varphi\nonumber\\
&& -6\left({2\over 5}\right)^{-{19\over 5}}g_s q^{-{2\over 5}}z^{9\over  
5}(\partial_0 a_z -\partial_z a_0) +{\lambda_S\over z^2}\left\{ {6\over 7}b^0_0 +{6\over 7}b^z_z  
+{3\over 4}b^i_i
-{32\over 49}\varphi\right\}=0,\qquad\ell\ge 0.
\label{ds}
\EQN
  From the metric equation (\ref{h}) we have the following
7 equations
\EQA
&& \hspace{-1.4cm} -{45\over 8}{1\over z}\partial_z b^z_z +{49\over  
4}{1\over z^2}b^0_0+28{1\over z^2}b^z_z -{25\over 8}\partial_z\partial_z  
b^i_i +{5\over 2}{1\over z}\partial_z b^i_i -{7\over 2} {1\over z^2}  
b^i_i+{75\over 14}\partial_z\partial_z \varphi -{135\over 14}{1\over  
z}\partial_z \varphi -21{1\over z^2} \varphi\nonumber\\
&& 
+{7\over 2}\left({2\over 5}\right)^{-{19\over 5}}g_s q^{-{2\over  
5}}z^{9\over 5}(\partial_0 a_z -\partial_z a_0) +{\lambda_S\over z^2}\left\{ -{1\over 2}b^z_z  
-{7\over 16}b^i_i +{6\over 7}\varphi\right\}=0,\qquad\ell\ge 0 ,
\label{h00s}
\EQN
\EQA
&& \hspace{-0.5cm} -{45\over 4}{1\over z}\partial_0 b^0_z +{45\over  
8}{1\over z}\partial_z b^0_0 +{49\over 4}{1\over z^2}b^0_0+28{1\over  
z^2}b^z_z +{25\over 8}\partial_0\partial_0 b^i_i +{35\over 4}{1\over  
z}\partial_z b^i_i -{7\over 2} {1\over z^2} b^i_i\nonumber\\
&&-{75\over 14}\partial_0\partial_0 \varphi -{135\over  
14}{1\over z}\partial_z \varphi -21{1\over z^2} \varphi +{7\over  
2}\left({2\over 5}\right)^{-{19\over 5}}g_s q^{-{2\over 5}}z^{9\over  
5}(\partial_0 a_z -\partial_z a_0)  \nonumber\\
&&\hspace{5cm}+{\lambda_S\over z^2}\left\{ -{1\over 2}b^0_0 -{7\over  
16}b^i_i +{6\over 7}\varphi\right\} =0,\qquad\ell\ge 0,
\label{hzzs}
\EQN
\EQA
&&  -{45\over 8}{1\over z}\partial_0 b^z_z -{25\over 8}\partial_z\partial_0  
b^i_i -{25\over 8}{1\over z}\partial_0 b^i_i    +{75\over  
14}\partial_0\partial_z \varphi +{\lambda_S\over z^2}\left\{ {1\over  
2}b^0_z\right\}=0,\qquad\ell\ge 0, \label{h0zs}
\EQN
\EQA
&& \hspace{-1cm} -25\partial_z\partial_z b^0_0 +70{1\over z}\partial_z b^0_0  
-98 {1\over z^2} b^0_0  +25\partial_0\partial_0 b^z_z -70{1\over  
z}\partial_z b^z_z +168 {1\over z^2} b^z_z +50\partial_0\partial_z b^0_z  
-140{1\over z}\partial_0 b^0_z\nonumber\\
&&\hspace{-1.4cm}+{175\over 8}\partial_0\partial_0 b^i_i -{175\over  
8}\partial_z\partial_z b^i_i +{315\over 8}{1\over z}\partial_z b^i_i -21  
{1\over z^2} b^i_i -{300\over 7}\partial_0\partial_0 \varphi +{300\over  
7}\partial_z\partial_z \varphi -120{1\over z}\partial_z \varphi +168 {1\over  
z^2} \varphi\nonumber\\
&& \hspace{-1.4cm}-28\left({2\over 5}\right)^{-{19\over 5}}g_s q^{-{2\over 5}}z^{9\over  
5}(\partial_0 a_z -\partial_z a_0)   +{\lambda_S\over z^2}\left\{ -{7\over 2}b^0_0 -{7\over  
2}b^z_z -{21\over 8}b^i_i
+6\varphi\right\} =0,\qquad\ell\ge 0 ,
\label{hiis}
\EQN
\EQA
&& \hspace{-1cm} -{25\over 8}\partial_z b^0_z -{25\over 8}\partial_0 b^z_z
- {175\over 64}\partial_0 {b^i}_i
+{95\over 8}{1\over z}b^0_z   +
{75\over 14}\partial_0\varphi +{7\over 2}\left({2\over 5}\right)^{-{19\over  
5}}g_s q^{-{2\over 5}}z^{9\over 5}
a_z =0 ,\qquad\ell\ge 1,
\label{h0is}\nonumber\\
\EQN
\EQA
&& -{25\over 8}\partial_0 b^0_z +{25\over 8}\partial_z  
b^0_0
+{175\over 64}\partial_z {b^i}_i
-{25\over 8}{1\over z}b^0_0 +{35\over 4}{1\over z}b^z_z     
-{75\over 14}\partial_z\varphi +
{75\over 14}{1\over z}\varphi\nonumber\\
&&\hspace{6cm}-{7\over 2}\left({2\over  
5}\right)^{-{19\over 5}}g_s q^{-{2\over 5}}z^{9\over 5}a_0 =0 ,\qquad\ell\ge 1,
\label{hzis}
\nonumber\\
\EQN
\EQA
&&  {1\over 2}b^0_0+{1\over 2}b^z_z+{3\over 8}{b^i}_i -{6\over 7}\varphi =0 
,\qquad\ell\ge 2 .
\label{hijs}
\EQN
 Finally,  the RR 1-form equation (\ref{a}) gives the following
3 equations
\EQA
&& \hspace{-2cm}  -{175\over 8}\partial_z b^0_0 -{175\over 8}\partial_z  
b^z_z +{175\over 8}\partial_z b^i_i +{75\over 2}\partial_z \varphi    
-\left({2\over 5}\right)^{-{19\over 5}}g_s q^{-{2\over 5}}z^{19\over  
5}\{{25\over 4}\partial_z(\partial_0 a_z -\partial_z a_0) \nonumber \\
&& \hspace{4cm} +{95\over 4}{1\over  
z}(\partial_0 a_z -\partial_z a_0) -{\lambda_S\over z^2}a_0 \} =0,\qquad\ell\ge 0,
\label{a0s}
\EQN
\EQA
&& \hspace{-1cm} -{175\over 8}\partial_0 b^0_0 -{175\over 8}\partial_0 b^z_z  
+{175\over 8}\partial_0 b^i_i +{75\over 2}\partial_0 \varphi\nonumber\\
&&  -\left({2\over  
5}\right)^{-{19\over 5}}g_s q^{-{2\over 5}}z^{19\over 5}\left\{{25\over  
4}\partial_0(\partial_0 a_z -\partial_z a_0) -{\lambda_S\over z^2}a_z \right\} =0,\qquad\ell\ge0,
\label{azs}
\EQN
\EQA
&&  -{25\over 4}\partial_0 a_0+{25\over 4}\partial_z a_z
+{45\over 4}{1\over z} a_z =0,\qquad\ell\ge 1.
\label{ais}
\EQN
In these equations, $\lambda_T, \lambda_V$ and
$\lambda_S$ are the eigenvalues of the
scalar Laplacian on $S^8$ on the tensor, vector
and scalar harmonics respectively, 
\EQ
\lambda_a =-\ell(\ell+7) + n_a ,
\EN
with $n_T=2, n_V=1$ and $ n_S=0$.

Let us now analyze the spectrum using these equations.
First, we treat the symmetric traceless tensor mode $b$. The
equation is already diagonalized as given in (\ref{symmtraceless}) which is  
solved by modified 
Bessel functions. 
The solution in the Euclidean space-time $(t\rightarrow -i\tau,\omega_{M}\rightarrow i\omega)$ which is regular at the origin is
\EQ
b=z^{{7\over 5}}K_{{1\over 5}(2\ell+7)}(\omega z) \,
\e^{-i\omega \tau}\qquad
(\ell =2,3,\ldots) .
\EN

For the vector modes, we shall first analyze the special case $\ell=1$.
We have 3 equations (\ref{h0iv}), (\ref{hziv}) and (\ref{aiv}) for 3 variables
$b^0,b^z,a$. However, there is a residual gauge symmetry corresponding to the 
Killing vector on  $S^8$
\[
\xi_0=\xi_z=0, \quad 
\xi_i= \zeta Y_i^{(\ell=1)}, 
\]
by  which the fields are transformed as 
\[
\delta b^0=-z^2\partial_0\zeta,\quad
\delta b^z=z^2\partial_z\zeta,\quad
\delta a=0.
\]
One of the three variables can be gauged away using this residual symmetry 
and another one is determined by the constraint equation,
so we expect one physical degree of freedom.
By examining the equations (\ref{h0iv}) and (\ref{hziv}), we  
first note that they are rewritten as
\[
\partial_z\left\{\hat{b}+7\hat{a}\right\} =0 , \quad 
\partial_0\left\{\hat{b}+7\hat{a}\right\} =0 ,
\]
where  $\hat{a}$ and $\hat{b}$ are defined by
\EQA
&&\hat{a} =g_s a ,\\
&&\hat{b} =-q^{2\over 5}z^{-{9\over 5}}\left({2\over 5}\right)^{9\over 5}
\left\{ \partial_0b^z +\partial_z b^0 -{2\over z}b^0\right\}.
\EQN
Thus, we can set
\EQ
\hat{b}+7\hat{a}=C\mbox{ (constant)}.
\label{constv0}
\EN
Using (\ref{constv0}) we can eliminate
$\hat{b}$ from (\ref{aiv}) and derive
\EQ
\partial_z\partial_z \hat{a} -\partial_0\partial_0 \hat{a}
+{9\over 5}{1\over z}\partial_z \hat{a} +{252\over 25}
{1\over z^2} (\hat{a} +C) =0
\EN
which is solved as 
\EQ
\hat{a}=z^{-{2\over 5}}K_{16\over 5}(\omega z)
\, \e^{-i\omega \tau}.
\EN
We have set the constant $C=0$ by invoking the 
boundary condition  
(see section 4) that 
the action evaluated with the solution have no contribution 
from the boundary at the origin $(z\rightarrow \infty)$.
This condition which we always assume in 
the present work essentially amounts to requiring that the 
self-adjointness of the kinetic operator  is 
not violated 
at the core, {\it i.e}, inside boundary $z\rightarrow \infty$. 
That is in general the natural boundary condition 
at points where the classical solutions are singular. 
For  such an analysis of linear 
perturbations around singular classical solutions, 
we  would like to refer the reader to \cite{ty}.   

 Now for the generic case $\ell\ge 2$,
there are four equations for three variables.
First it is easy to check that the set of the equations are consistent by  
showing
that (\ref{hijv}) is derived from (\ref{h0iv}), (\ref{hziv}) and
(\ref{aiv}).  
Since two of them (\ref{hziv}) and (\ref{aiv}) contain time derivative of 
second order, we note that there are two physical degrees of freedom. 
To obtain the diagonalized
excitations, we make a linear combination
of the derivatives of (\ref{h0iv}) and (\ref{hziv}) which reads
\EQ
\partial_z\partial_z (\hat{b}+7\hat{a})-\partial_0\partial_0  
(\hat{b}+7\hat{a})+{9\over 5}{1\over z}\partial_z (\hat{b}+7\hat{a})
+{4\over 25}\{-\ell(\ell +7) +8\}{1\over z^2}\hat{b} =0.
\label{vector1}
\EN
The coupled equations (\ref{aiv}) and (\ref{vector1}) can be diagonalized
by defining
\EQA
&&\hat{a}_1=\hat{a}-(\ell+1)\hat{b} ,\\
&&\hat{a}_2=\hat{a}+{1\over \ell +6}\hat{b} ,
\EQN
and the solutions are given by
\EQA
&&\hat{a}_1=z^{-{2\over 5}}K_{{2\over 5}(\ell+7)}(\omega z)\, \e^{-i\omega  
\tau} ,\qquad
(\ell =1,2,\ldots)\\
&&\hat{a}_2=z^{-{2\over 5}}K_{{2\over 5}\ell }(\omega z)\, \e^{-i\omega  
\tau} ,\qquad
(\ell =2,3,\ldots)
\EQN 
where we have included also the $\ell=1$ mode.

Let us next turn to the scalar modes.
We shall start by examining the modes
with lowest angular momentum $\ell=0$. There are 7 equations for 7 variables
$\varphi, b^0_0, b^0_z, b^z_z, b^i_i, a_0$ and $a_z$. As in the vector case,
we have to examine the residual gauge symmetry for the equations.
There are two kinds of residual gauge symmetries. One is the two dimensional
diffeomorphism for which the gauge parameter is of the form
\[
\xi_0 =\zeta_0 Y^{(\ell=0)}, \quad 
\xi_z =\zeta_z Y^{(\ell=0)}, \quad
\xi_i =0,
\]
which leads to 
\EQA
&&\delta b^0_0=2\partial_0\zeta^0 -2{1\over z}\zeta^z,\quad
\delta b^0_z=-\partial_0\zeta^z +\partial_z\zeta^0,\quad
\delta b^z_z=2\partial_z\zeta^z -2{1\over z}\zeta^z,\quad
\delta b^i_i=0,\quad\nonumber\\
&&\delta \varphi ={21\over 10}{1\over z}\zeta^z,\quad
\delta a_0=-{1\over g_s}({2\over 5})^{14\over 5}q^{2\over 5}
z^{-{14\over 5}}\{\partial_0\zeta^0 -{14\over 5}{1\over z}\zeta^z\},\quad
\delta a_z=-{1\over g_s}({2\over 5})^{14\over 5}q^{2\over 5}
z^{-{14\over 5}}\partial_z\zeta^0. \nonumber
\EQN
The other is the RR 1-form gauge transformation
\[
\Lambda=\lambda Y^{(\ell=0)},
\]
which leads to
\[
\delta a_0=\partial_0\lambda,\quad
\delta a_z=\partial_z\lambda.
\]
As in the vector case, 3 variables can be gauged away using this gauge 
freedom and another 3
variables will be determined by the constraints. 
We thus expect only one physical degree of
freedom.
For $\ell=0$, we note that (\ref{a0s}) and (\ref{azs}) can be
rewritten, respectively, as
\EQ
\partial_z\left\{ -{7\over 2}b^0_0 -{7\over 2}b^z_z +{7\over 2}b^i_i  
+6\varphi
-\left({2\over 5}\right)^{-{19\over 5}}g_s q^{-{2\over 5}}z^{19\over  
5}(\partial_0 a_z -\partial_z a_0)\right\}=0 ,
\EN
\EQ
\partial_0\left\{ -{7\over 2}b^0_0 -{7\over 2}b^z_z +{7\over 2}b^i_i  
+6\varphi
-\left({2\over 5}\right)^{-{19\over 5}}g_s q^{-{2\over 5}}z^{19\over  
5}(\partial_0 a_z -\partial_z a_0)\right\}=0. 
\EN
Therefore we can set the quantity in the parenthesis to be
a constant ($C_1$)
\EQ
-{7\over 2}b^0_0 -{7\over 2}b^z_z +{7\over 2}b^i_i +6\varphi
-\left({2\over 5}\right)^{-{19\over 5}}g_s q^{-{2\over 5}}z^{19\over  
5}(\partial_0 a_z -\partial_z a_0) = C_1
\label{const01}
\EN
which enables us to eliminate $(\partial_0 a_z -\partial_z a_0)$.
(\ref{h0zs}) can now be written as
\EQ
\partial_0\left\{ -{9\over 4}{1\over z} b^z_z -{5\over 4}\partial_z b^i_i  
-{5\over 4}{1\over z} b^i_i +{15\over 7}\partial_z \varphi \right\}=0
\EN
and, using (\ref{const01}), (\ref{h00s}) can be written as
\EQ
\partial_z\left\{ z^{-{9\over 5}}\left( -{9\over 4}{1\over z} b^z_z -{5\over  
4}\partial_z b^i_i -{5\over 4}{1\over z} b^i_i +{15\over 7}\partial_z  
\varphi \right)\right\}=0 .
\EN
 Thus we have
\EQ
-{9\over 4}{1\over z} b^z_z -{5\over 4}\partial_z b^i_i -{5\over 4}{1\over  
z} b^i_i +{15\over 7}\partial_z \varphi =C_2 z^{9\over 5}  .
\label{const02}
\EN
We can set the constants $C_1=C_2=0$ for the same reason as in the vector case.
Combining the equations (\ref{ds}), (\ref{hzzs}), (\ref{hiis}) and
using the constraints (\ref{const01}), (\ref{const02}), we obtain the 
diagonalized equation for a physical mode $b^i_i$
\EQ
 -\partial_0\partial_0 b^i_i + \partial_z\partial_z b^i_i -{9\over 5}{1\over  
z}\partial_z
 b^i_i -{392\over 25}{1\over z^2}b^i_i=0.
\label{biieqofmotion}
\EN
All the other variables can be gauged away or determined by
the constraints. Indeed, if we adopt the gauge conditions  $b^0_0=b^z_z=a_0=0$,
the rest of the variables $a^z, b^0_z,\varphi$ are determined by (\ref{const01}), (\ref{const02}), and another constraint equation which follow from 
(\ref{ds}), (\ref{hzzs}), (\ref{hiis}), (\ref{const01}) and (\ref{const02})
\[
{5\over 2}\partial_z\partial_0 b^0_z -{5\over 2}{1\over z}\partial_0 b^0_z
-{5\over 4}{1\over z}\partial_z b^i_i -{19\over 2}{1\over z^2}b^i_i =0 .
\]
It is easy to 
check the consistency of the result by seeing that the set of the 
solutions obtained
above actually satisfies the original equations.

We can repeat the similar analysis for $\ell=1$.
In this case, there are 10 equations for 7 variables. We first check that
3 of them are consequences of the other equations: for $\ell\ge 1,
(\ref{h00s}) + (\ref{h0zs})+(\ref{azs})  \rightarrow
(\ref{h0is}),
(\ref{h00s})+(\ref{hzzs})+(\ref{h0zs})+(\ref{ds})
+(\ref{a0s}) \rightarrow (\ref{hzis}) ,
(\ref{a0s}) + (\ref{azs}) \rightarrow (\ref{ais})$.
There is a residual gauge symmetry associated with the 
conformal Killing vector on S$^8$, (
the so-called 'conformal diffeomorphism' \cite{krn}). 
It is the diffeomorphism
whose parameters take the form
\[
\xi_0=\partial_0 \eta Y^{(\ell=1)}, \quad 
\xi_z=\partial_z \eta Y^{(\ell=1)}, \quad 
\xi_i=-\eta D_i Y^{(\ell=1)}.
\]
which leads to
\EQA
&&\delta b^0_0=-z^2\partial_0\partial_0\eta -z\partial_z\eta,\quad
\delta b^0_z=-z^2\partial_0\partial_z\eta -z\partial_0\eta,\quad
\delta b^z_z=z^2\partial_z\partial_z\eta +z\partial_z\eta,\quad
\delta b^i_i={32\over 25}\eta\nonumber\\
&&\delta\varphi={84\over 5}z\partial_z\eta,\quad
\delta a_0={7\over 5}{1\over g_s}({2\over 5})^{14\over 5}q^{2\over 5}
z^{-{14\over 5}}z\partial_z\eta,\quad
\delta a_z={7\over 5}{1\over g_s}({2\over 5})^{14\over 5}q^{2\over 5}
z^{-{14\over 5}}z\partial_0\eta .
\EQN
We find two physical degrees of freedom for $\ell=1 $ mode.
Since the final result, however,  is more conveniently
summarized by using the dynamical modes
for generic $\ell$, let us now turn to the case $\ell\ge 2$.

First, there are 11
equations for 7 variables. We can check that
4 of them are consequences of the
other equations. In addition to the relations mentioned in
the $\ell=1$ case above, we have 
$(\ref{h0is})+(\ref{hzis})+(\ref{hiis}) \rightarrow
(\ref{hijs})$ for $\ell\neq 1$. 
Only 3 of the remaining 7 equations are the equations of motion.
The combinations of the 3 dynamical
components which diagonalize the equations of
motion are given as
\[
s_1= z^{-7/5}\Bigl(-{7(-\ell+ 7)\over 16}b_i^i + f\Bigr) ,
\]
\EQ
s_2=z^{-12/5}
\Bigl(
{\ell(\ell+7)\over 4}b^0_z +{5\ell(\ell+7)\over 21}z\partial_0\varphi
+{5\{-\ell(\ell+7)+49\}\over 64}z\partial_0b^i_i
-{5\over 8}z\partial_0 f
\Bigr) ,
\EN
\[
s_3=z^{-7/5}\Bigl(-{7(\ell +14)\over 16}b_i^i +f \Bigr) ,
\]
where
\EQ
f= g_s\bigl({5\over 2}\bigr)^{19/5}q^{-2/5}z^{19/5}
(\partial_0a_z-\partial_za_0) .
\EN
It can be checked that $s_3$ is a gauge mode for $\ell=1$. 
That is, we can proceed in the same way as $\ell\ge 2$ case if we impose (\ref{hijs}) 
using the residual gauge symmetry of the  $\ell=1$ mode. However,  
 there still remains a gauge symmetry which enables us to
gauge away $s_3$. 
Also, note that the result agrees with the analysis for $\ell=0$. 
In this case, we impose (\ref{h0is}), 
(\ref{hzis}), (\ref{ais}) and (\ref{hijs}) using the residual gauge symmetries.
From the constraint (\ref{const01}) and the gauge condition (\ref{hijs}),
we can see that $s_2$ and $s_3$ vanish for $\ell=0$.
The final results for the equations and their solutions are summarized as
\[
 -\partial_0\partial_0  s_1+ \partial_z\partial_z  s_1 +{1\over z}\partial_z
 s_1 -{(2\ell+21)^2\over 25}{1\over z^2}  s_1=0,
\]
\EQ
 -\partial_0\partial_0 s_2 + \partial_z\partial_z s_2  +
{1\over z}\partial_z
 s_2-{(2\ell+7)^2\over 25}{1\over z^2}s_2=0 ,
\EN
\[
 -\partial_0\partial_0   s_3 + \partial_z\partial_z   s_3
+ {1\over z}\partial_z
   s_3 -{(2\ell-7)^2\over 25}{1\over z^2}  s_3=0,
\]
\EQA
&&s_1=K_{{1\over 5}(2\ell +21)}(\omega z)\e^{-i\omega  
\tau}\qquad\mbox{for}\quad\ell\ge 0  
,\nonumber\\
&&s_2=K_{{1\over 5}(2\ell +7)}(\omega z)\e^{-i\omega  
\tau}\qquad\mbox{for}\quad\ell\ge 1 ,  
\\
&&s_3=K_{{1\over 5}(2\ell -7)}(\omega z)\e^{-i\omega  
\tau}\qquad\mbox{for}\quad\ell\ge 2.
\nonumber
\EQN

\subsection{The fluctuations of NS-NS 2 form and 
RR 3-form fields}

 Let us next study the fluctuations whose background fields 
are trivial; 
$B^{(total)}_{\mu\nu}=\hat{B}_{\mu\nu}$ and $A^{(total)}_{\mu\nu\rho}=\hat{A}_{\mu
\nu\rho}$.  The equations of motions for NSNS 2 form and 
RR 3 form fields are 
\EQ
D^\sigma\left( \e^{-{10\over 7}\tilde{\phi}}H_{\sigma\mu\nu}
\right)
+g_s^2 D^\sigma\left( A^\rho\e^{{2\over 7}\tilde{\phi}}\tilde{F}_{\sigma
\rho\mu\nu}\right)=0 ,
\label{b}
\EN
\EQ
D^\sigma\left( \e^{{2\over 7}\tilde{\phi}}\tilde{F}_{
\sigma\mu\nu\rho}\right)=0 , 
\label{a3}
\EN
respectively. 
The natural gauge conditions for performing 
the harmonic analysis are 
\EQA
&&D^i \hat{B}_{0i} =D^i \hat{B}_{zi}=D^i \hat{B}_{ij}=0\nonumber\\
&&D^i \hat{A}_{0zi} =D^i \hat{A}_{0ij}=D^i \hat{A}_{zij}=
D^i \hat{A}_{ijk}=0 
\label{gaugecond2}
\EQN
which make us possible to expand as 
\EQA
&&\hat{B}_{0z}=\beta_{0z}Y,\quad\hat{B}_{0i}=\beta_{0}Y_i,
\quad\hat{B}_{zi}=\beta_{z}Y_i,\quad\hat{B}_{ij}=\beta Y_{[ij]}
\nonumber\\
&&\hat{A}_{0zi}=\alpha_{0z}Y_i,\quad\hat{A}_{0ij}=\alpha_{0}Y_{[ij]},
\quad\hat{A}_{zij}=\alpha_{z}Y_{[ij]},\quad
\hat{A}_{ijk}=\alpha Y_{[ijk]}
\EQN
where $Y_i$, $Y_{[ij]}$ and $Y_{[ijk]}$ are called 1-form, 2-form
and 3-form harmonics respectively whose definitions and 
properties are summarized in Appendix.  For notational 
brevity, we abbreviated the sum over $\ell$. 

We first comment on 
the counting 
of the degrees of freedom. 
The number of the fields are originally 45 ($\hat{B}_{\mu\nu}$) +
120 ($\hat{A}_{\mu\nu\rho}$). There are 8  ($\hat{B}_{\mu\nu}$) +
28 ($\hat{A}_{\mu\nu\rho}$) constraints coming from the field equations, 
and the gauge conditions (\ref{gaugecond2}) eliminate 9+36 components.
 Thus we have 84 physical components.
In terms of the expansion by spherical harmonics, 
we will see that  for the 
generic case ($\ell\ge1$) 
there is no radial degree of freedom for the scalar, one for 1-form,
two for divergenceless 2-form and one for divergenceless 3-form, 
which indeed sum up to 84 degrees of freedom.

As in the previous subsection, let us start from the case of 
lowest angular momentum.  
From the definition, there is no $\ell=0$ contribution 
for $p$-form harmonics with nonzero $p$. Therefore, only possibility 
for $\ell=0$ mode is the $S^8$ scalar component $\beta_{0z}$. 
However, it is easy to see that the scalar component 
$\beta_{0z}$ can be gauged away using the residual gauge symmetry for $\hat{B}_{\mu\nu}$ 
with parameter
\[
\Lambda_0=\lambda_0 Y^{(\ell=0)},\quad\Lambda_z=\lambda_z Y^{(\ell=0)}
\quad\Lambda_i=0.
\]
which leads to
\[
\delta \beta_{0z}=\partial_0\lambda_z -\partial_z\lambda_0.
\]
Thus there is no $\ell=0$ mode in the physical spectrum. 

Now, we shall analyse the generic $\ell\ge 1$ case.
The equations listed in the rest of this section are 
valid for $\ell\ge 1$.
The component equations for the scalar harmonics are
\EQA
&&\beta_{0z}=0,\\
&&\partial_{z}\beta_{0z}-{1\over z}\beta_{0z}=0,\\
&&\partial_{0}\beta_{0z}=0,
\EQN
which are solved trivially by 
\EQ
\beta_{0z}=0.
\EN
The equations for 1-form harmonics are 
\EQA
&&\partial_z\partial_z\beta_0-\partial_z\partial_0\beta_z
-{1\over z}\partial_z\beta_0+{1\over z}\partial_0\beta_z
-{4\over 25}(\ell+6)(\ell+1){1\over z^2}\beta_0=0 ,
\label{oneform1}\\
&&\partial_z\partial_0\beta_0-\partial_0\partial_0\beta_z
-{4\over 25}(\ell+6)(\ell+1){1\over z^2}\hat{\alpha}_{0z}
=0 ,
\label{oneform2}\\
&&\partial_0\beta_0 -
\partial_z\hat{\alpha}_{0z}+3{1\over z}\hat{\alpha}_{0z}=0 ,
\label{oneform3}\\
&&\hat{\alpha}_{0z}-\beta_z=0 ,
\label{oneform4}\\
&&\partial_z(\hat{\alpha}_{0z}-\beta_z)-{1\over 5}{1\over z}
(\hat{\alpha}_{0z}-\beta_z)=0 ,
\label{oneform5}\\
&&\partial_0(\hat{\alpha}_{0z}-\beta_z)=0 ,
\label{oneform6}
\EQN
where we have defined
\EQ
\hat{\alpha}_{0z}= g_s\left({2\over 5}\right)^{-{14\over 5}}
q^{-{2\over 5}}z^{14\over 5}\alpha_{0z}.
\EN
There are six equations for three variables.
The consistency of the equations is proved by noticing
that (\ref{oneform3}) is derived by (\ref{oneform1}) 
and (\ref{oneform2}), and that (\ref{oneform5}) 
and (\ref{oneform6}) are consequences of (\ref{oneform4}).
Using the similar arguments 
as in the previous subsection, this shows that there is only one physical degree of freedom. The radial equation is 
obtained by combining $z$-derivative of (\ref{oneform1}) and
0-derivative of (\ref{oneform2}) 
\EQ
\partial_z\partial_z u +{1\over z}\partial_z u
-\partial_0\partial_0 u -{(2\ell +7)^2\over 25}{1\over z^2}
u=0 ,
\EN
where 
\EQ
u=\partial_z\beta_0 -\partial_0\beta_z.
\EN
The solution is
\EQ
u=K_{{1\over 5}(2\ell+7)}(\omega z)\,\e^{-i\omega \tau} .
\EN

For the 2-form modes, there are four equations for 
three variables $\beta$, $\alpha_0$ and $\alpha_z$.
\EQA
&&\hspace{-0.8cm} -\partial_0\partial_0 \hat{\beta} +\partial_z(\partial_z\alpha_0
-\partial_0\alpha_z)-{1\over 5}{1\over z}(\partial_z\alpha_0
-\partial_0\alpha_z)-{4\over 25}(\ell+5)(\ell+2){1\over z^2}
\alpha_0 =0 ,
\label{twoform1}\\
&&\hspace{-0.5cm} -\partial_z\partial_z \hat{\beta} -{27\over 5}{1\over z}
\partial_z \hat{\beta}+{4\over 25}(\ell+9)(\ell-2){1\over z^2}
\hat{\beta}\nonumber\\
&&+\partial_z(\partial_z\alpha_0
-\partial_0\alpha_z)+{13\over 5}{1\over z}(\partial_z\alpha_0
-\partial_0\alpha_z)-{4\over 25}(\ell+5)(\ell+2){1\over z^2}
\alpha_0=0 ,
\label{twoform2}\\
&&-\partial_0\partial_z \hat{\beta}-{14\over 5}{1\over z}
\partial_0\hat{\beta}
+\partial_0(\partial_z\alpha_0 -\partial_0\alpha_z)
-{4\over 25}(\ell+5)(\ell+2){1\over z^2}
\alpha_z=0 ,
\label{twoform3}\\
&&\hspace{1cm} -\partial_0\hat{\beta}+\partial_0\alpha_0
-\partial_z\alpha_z-{3\over 5}{1\over z}\alpha_z=0 ,
\label{twoform4}
\EQN
where  
\EQ
\hat{\beta}={1\over g_s}\left({2\over 5}\right)^{14\over 5}
q^{2\over 5}z^{-{14\over 5}}\beta.
\EN
The consistency of the equations can be proved by checking that
(\ref{twoform4}) is derived from (\ref{twoform2}) and 
(\ref{twoform3}).  There are two physical components. 

To solve the equations, we define the combination
$\hat{\alpha}_1=z(\partial_z\alpha_0 -\partial_0\alpha_z
-\partial_z\hat{\beta})$.
   From (\ref{twoform1}) and (\ref{twoform2}), we find
\EQ
-\partial_0\partial_0 \hat{\beta}+\partial_z\partial_z
 \hat{\beta}+{13\over 5}{1\over z}\partial_z\hat{\beta}
-{4\over 25}(\ell+9)(\ell-2){1\over z^2}\hat{\beta}
-{14\over 5}{1\over z^2}\hat{\alpha}_1=0
\label{detwoform1}
\EN
and, from (\ref{twoform2}) and (\ref{twoform3}), 
\EQ
-\partial_0\partial_0 \hat{\alpha}_1+\partial_z\partial_z
 \hat{\alpha}_1+{13\over 5}{1\over z}\partial_z\hat{\alpha}_1
-{4\over 25}(\ell+5)(\ell+2){1\over z^2}\hat{\alpha}_1
-{14\over 5}\left\{-\partial_0\partial_0 \hat{\beta}+\partial_z
\partial_z\hat{\beta}+{13\over 5}{1\over z}\partial_z\hat{\beta}
\right\}
=0 .
\label{detwoform2}
\EN
The coupled equations (\ref{detwoform1})
and (\ref{detwoform2}) can be diagonalized by defining
\EQA
&&v_1=z^{4\over 5}\left(\hat{\alpha} -{2\over 5}(\ell+9)\hat{\beta}_1 ,
\right)\\
&&v_2=z^{4\over 5}\left(\hat{\alpha} +{2\over 5}(\ell-2)\hat{\beta}_1 ,
\right)\EQN
which lead to 
\EQA
&&-\partial_0\partial_0 v_1+\partial_z\partial_z
 v_1+{1\over z}\partial_z v_1
-{4\over 25}\ell^2{1\over z^2}v_1=0 ,\\
&&-\partial_0\partial_0 v_2+\partial_z\partial_z
 v_2+{1\over z}\partial_z v_2
-{4\over 25}(\ell+7)^2{1\over z^2}v_2=0.
\EQN
The solutions are 
\EQA
&&v_1=K_{{2\over 5}\ell}(\omega z)\,\e^{-i\omega \tau} ,\\
&&v_2=K_{{2\over 5}(\ell+7)}(\omega z)\,\e^{-i\omega \tau} .
\EQN

Finally, for 3-form modes, there is only 
one equation for one variable
\EQ
-\partial_0\partial_0 \alpha+\partial_z\partial_z
 \alpha+{3\over 5}{1\over z}\partial_z \alpha
-{4\over 25}(\ell+4)(\ell+3){1\over z^2}\alpha=0 ,\\
\EN
which is solved as
\EQ
\alpha=z^{1\over 5}K_{{1\over 5}(2\ell+7)}(\omega z)\,\e^{-i\omega 
\tau}.
\EN
We have completed the harmonic analysis of bosonic 
fluctuations around the D0 background.

\section{Supergravity-Matrix theory correspondence} 
\setcounter{equation}{0} 

We now try to establish the correspondence between 
10 dimensional supergravity and Matrix theory in the large $N$ limit 
using the results of the harmonic analysis. 
We will follow the specific prescription in Euclidean metric given in 
\cite{gpk} by assuming that the `boundary' on the side 
of supergravity is located at the limit of the region 
of validity for the near horizon condition, 
namely, at $z=q^{1\over 7} \, \, \, ( r\propto q^{1/7}) $. Remember, as emphasized 
in section 2,  that this is compulsory because we are 
interested in the region $g_sN > 1$.  
Since the perturbing fields are diagonalized, we can 
discuss each diagonalized components separately.  

First we consider the traceless-symmetric tensor mode 
as a simple example and then present the general result. 
The relevant part of
the action is
\[
\hspace{-2cm} S={1\over 8\kappa^2}\int d\Omega_8 Y^I_{ij}Y^J_{i'j'}g^{ii'}g^{jj'}
\int d\tau
\int dz \left( {2\over 5}\right)^{9\over 5}q^{7\over 5}z^{-{9\over 5}}
[
\partial_z b^I\partial_z b^J+\partial_0 b^I\partial_0 b^J 
\]
\EQ
\hspace{5cm} +{4\over 25}\ell(\ell+7)
{1\over z^2}b^I b^J ]
\EN
where we have introduced the indices $I,J$, labeling  the harmonics
that have been suppressed in the last section.
The action evaluated for the classical solution which we call $K$ can 
be expressed in terms of the boundary value of the field, 
\EQ
K={1\over 8\kappa^2}C \delta^{IJ}\int d\tau
 \left( {2\over 5}\right)^{9\over 5}q^{7\over 5}
\left[ z^{-{9\over 5}}b^I\partial_z b^J\right]_{q^{1\over 7}}^{\infty}.
\EN
The harmonics are normalized as
$\int d\Omega_8 Y^I_{ij}Y^J_{i'j'}g^{ii'}g^{jj'}=C\delta^{IJ}$ 
with $C$ being a numerical constant independent of $I, J$ and of 
$q\propto g_sN$. In what follows we always suppress the 
indices $I$. 
The solution $b$ satisfying the boundary condition
$b(q^{1/7},\tau)=\int d\omega \e^{-i\omega\tau}f_{\omega}$
is 
\EQ
b(\tau,z)=\int d\omega \e^{-i\omega\tau}\tilde{b}_\omega(z)
f_{\omega}
\EN
where $\tilde{b}_{\omega}(z)$ is the solution of the radial equation
normalized to 1 at the boundary $(z\rightarrow   q^{1\over 7})$, 
\EQ
\tilde{b}^I_\omega(z)={z^{7\over 5}K_{{1\over 5}(2\ell+7)}(\omega z)
\over q^{1\over 5}K_{{1\over 5}(2\ell+7)}(\omega q^{1\over 7})}
\qquad (\mbox{for}\quad\ell=2,3,\ldots)  .
\EN
Then the action is evaluated as
\EQA
K&=&-{\pi\over 4\kappa^2}C\left({2\over 5}\right)^{9\over 5}q
\int d\omega f_{\omega} f_{-\omega}\left[ -{2\over 5}\ell +
({\rm terms \, \, analytic\, \,  in\, \, } \omega) 
\right.\nonumber\\
&&\left.\qquad +(q^{1\over 7}\omega)^{{2\over 5}(2\ell+7)}
\left\{ -2^{-{4\over 5}\ell -{9\over 5}}{\Gamma(-{2\over 5}\ell
-{2\over 5})\over \Gamma({2\over 5}\ell+{7\over 5})}+ ({\rm terms \, \, analytic\, \,  in\, \, } \omega)\right\}\right] .
\label{actionk}
\EQN
 The connected part of the two-point function of the operator of Matrix theory which 
couples to $b$ is now given by the second variation of (\ref{actionk})
with respect to $f_{\omega}$. 
The leading part   in the long-time region is 
\EQA
\left<{\cal O}(\tau){\cal O}(\tau ')\right>_c &=&
\int d\omega\int d\omega '\e^{i\omega \tau}\e^{i\omega' \tau'}
\left< {\cal O}(\omega){\cal O}(\omega ')\right>_c
=-\int d\omega\int d\omega '\e^{i\omega \tau}\e^{i\omega' \tau'}
{\delta\over \delta
f_{\omega}}{\delta\over \delta f_{\omega '}}
K[\lambda]\nonumber\\
&=&2^{-{2\over 5}(2\ell+7)}\pi^2{\Gamma({4\over 5}\ell
+{19\over 5})\over (\Gamma({2\over 5}\ell+{7\over 5}))^2}\left({2\over 5}
\right)^{9\over 5}
C{1\over \kappa^2}q^{{4\ell\over 35}+{7\over 5}}
{1\over \vert\tau -\tau'\vert^{{4\ell\over 5}+{19\over 5}}}
\label{tstensorcorrelator}
\EQN
where we have ignored the short time delta-function 
singularities coming from the 
analytic terms in $\omega$ as usual.\footnote{While we were preparing the 
present manuscript, 
we came to know that a similar result ($\ell=0$) for D1-brane   has been 
 given in a recent paper \cite{ahlp} for a certain particular mode of metric fluctuations. }
This result shows that the scaling dimension of the operator  under the 
generalized scaling transformation $ 
\tau \rightarrow \lambda^{-1} \tau,  g_s \rightarrow \lambda^3 g_s$ is 
\EQ
\Delta = 1+{4\over 7}\ell  .
\label{confdimen}
\EN
The final results of all the modes including this case can succinctly be 
 expressed in a unified manner as follows.  

The linearized action, up to a total derivative term 
which does not contribute to the large time 
behavior of the corrrelators,  is
\EQ
S={1\over 8\kappa^2}C \delta^{IJ}\int d\tau\int dz q z
\left[\partial_z g^I\partial_z g^J+\partial_0 g^I\partial_0 g^J 
+\nu^2{1\over z^2}g^I g^J\right]
\label{linearaction}
\EN
where $g^I$ denotes each diagonalized field and $C\delta^{IJ}$ comes 
from the normalization of each kind of the 
spherical harmonics.  The diagonalized fields $g^I$ are 
redefined by making suitable scalings by powers 
of $z$ such that they obey the Bessel equations 
without any pre-factor of $z$.  We also note that,  
in terms of $g^I$ the 
boundary condition at the inside boundary, $r=0 $ or $z\rightarrow \infty$, corresponding to the 
self-adjointness of the kinetic operator which enables us 
to extract the correlators from the near-horizon 
boundary, is $zg^I\partial_z g^I\rightarrow 0$. 
The constant $\nu$ is  the order of the Bessel function.  The leading part of the correlator, 
omitting numerical proportional constant,  is 
\EQ
\left<{\cal O}(\tau){\cal O}(\tau ')\right>_c=
{1\over \kappa^2}q^{1+{2\over 7}\nu}
{1\over \vert\tau -\tau'\vert^{2\nu +1}} ,
\label{2point}
\EN
giving  the general formula for the scaling dimension 
of the operator $\cal{O}$
\EQ
\Delta= -1 +{10\over 7}\nu  .
\EN
The results are summarized in the tables below. 
\begin{center}
{\bf Table 1}
\end{center}
\[
\begin{array}{|c||c|c|c|c|c|c|}
\hline
\mbox{SUGRA fields}&h^i_j&\multicolumn{2}{|c|}{h^0_i, h^z_i, 
\hat{A}_i}&\multicolumn{3}{|c|}{\hat{\phi},h^0_0,h^0_z,h^z_z,
h^i_i,\hat{A}_0,\hat{A}_z}\\
\hline\hline
\mbox{physical modes}&b&a_1&a_2&s_1&s_2&s_3\\
\hline
\mbox{order} \, \, \nu&{1\over 5}(2\ell+7)&{2\over 5}(\ell+7)&{2\over 5}\ell
&{1\over 5}(2\ell+21)&{1\over 5}(2\ell+7)&{1\over 5}(2\ell-7)\\
\hline
\mbox{dimensions of }{\cal O}&1+{4\over 7}\ell&3+{4\over 7}\ell&-1
+{4\over 7}\ell&5+{4\over 7}\ell&1+{4\over 7}\ell&
-3+{4\ell\over 7}\\
\hline
\mbox{regions of }\ell&\ell\ge2&\ell\ge1&\ell\ge2&\ell\ge0
&\ell\ge1&\ell\ge2\\
\hline
\end{array}
\]
\vspace{1cm}
\begin{center}
{\bf Table 2}
\end{center}
\[
\begin{array}{|c||c|c|c|c|}
\hline
\mbox{SUGRA fields}&\hat{B}_{0i},\hat{B}_{zi}&
\multicolumn{2}{|c|}{\hat{B}_{ij},\hat{A}_{0ij},\hat{A}_{zij}}&
\hat{A}_{ijk}\\
\hline\hline
\mbox{physical mode}&u&v_1&v_2&\alpha\\
\hline
\mbox{order} \, \, \nu&{1\over 5}(2\ell +7)&{2\over 5}(\ell +7)&{2\over 5}\ell&
{1\over 5}(2\ell +7)\\
\hline
\mbox{dimensions of }{\cal O}&1+{4\over 7}\ell&3+{4\over 7}\ell&
-1+{4\over 7}\ell&1+{4\over 7}\ell\\
\hline
\mbox{regions of }\ell&\ell\ge1&\ell\ge1&\ell\ge1&\ell\ge1\\
\hline
\end{array}
\]

\vspace{0.5cm}
We note that, 
up to a total derivative term,  the 
above linearlized action (\ref{linearaction}) 
is equivalent to the s-wave part of the following special 
(Euclidean) action for a massive scalar field $\psi$ after making 
a scaling transformation $\psi \rightarrow \exp({2\tilde{\phi}/3})\psi $, 
\EQ
S={1\over 2}\int d^{10}x\, \sqrt{-g} \, 
\e^{-2\phi} 
[g^{\mu\nu}\partial_{\mu}\psi\partial_{\nu}\psi 
+m^2 \e^{-2\phi/7}\psi^2]
\EN
where the mass is related to the order $\nu$ by 
\EQ
\tilde{m}^2 +{49\over 25}
=\nu^2  ,
\EN
with $25\tilde{m}^2/4=\Bigl({q\over g_s}\Bigr)^{2/7}m^2$. 
The total derivative term, being analytic in $\omega$, does not 
contribute to the long-time behavior of the correlation function. 
Thus the relation between the mass and the 
generalized conformal dimension is
\EQ
\Delta=-1 \pm {10\over 7}\sqrt{\tilde{m}^2 + {49\over 25}} .
\EN
For example, for the traceless-symmetric tensor mode, 
we have $\tilde{m}^2\rightarrow {4\ell(\ell+7)\over 25}$ 
by choosing the branch of the square root 
such that $\Delta$ has the positive coefficient with respect to $\ell$.  It is an interesting question whether the above mass 
can be related to the Casimir operator  in the representation of  the {\it generalized} conformal symmetry.  

We are now ready to discuss the correspondence of 
the spectrum of supergravity to the Matrix-theory 
operators.  Various currents (more precisely $x^-$-integrated currents) of Matrix theory have been 
identified in the work \cite{taylor} from the results of  perturbative calculations for the interactions 
between pairs of general background configurations 
of Matrix theory.  Let us quote their results below 
using their convention. 
We will only present the parts of the definitions to the 
extent that are needed in order to
 read off their generalized conformal dimensions.  
These operators have definite dimensions 
under the generalized conformal transformations. 
Note that the generalized conformal dimensions 
are {\it not} identical to the `engineering dimensions'. 
For full expressions of these operators, we refer the 
reader to \cite{taylor}. 
Anticipating the correspondence with 11 dimensional 
supergravity, we will use their notations 
using the 11 dimensional light-cone indices. 
Corresponding to 11 dimensional metric, we have
\EQA
T^{++}&=&{1\over R}{\rm STr}(1) ,\nonumber\\
T^{+i}&=&{1\over R}{\rm STr}(\dot{X}_i) , \nonumber\\
T^{+-}&=&{1\over R}{\rm STr}({1\over 2}\dot{X}_i
\dot{X}_i + \cdots) ,\nonumber\\
T^{ij}&=&{1\over R}{\rm STr}(\dot{X}_i\dot{X}_j
+\cdots) ,\label{energymomentum}\\
T^{-i}&=&{1\over R}{\rm STr}({1\over 2}\dot{X}_i
\dot{X}_j\dot{X}_j +\cdots), \nonumber\\
T^{--}&=&{1\over 4R}{\rm STr}(F_{ab}F_{bc}F_{cd}F_{da}+\cdots) .\nonumber
\EQN
Corresponding to 11 dimensional 3-form, we have
\EQA
J^{+ij}&=&{1\over 6R}{\rm STr}(F_{ij}) ,\nonumber\\
J^{+-i}&=&{1\over 6R}{\rm STr}(F_{ij}\dot{X}_j +\cdots) , \nonumber\\
J^{ijk}&=&{1\over 6R}{\rm STr}(-\dot{X}_iF_{jk} -
\dot{X}_jF_{ki}-\dot{X}_kF_{ij} +\cdots) , \\
J^{-ij}&=&{1\over 6R}{\rm STr}(\dot{X}_i\dot{X}_k
F_{kj} -\dot{X}_j\dot{X}_k
F_{ki}+ \cdots) .\nonumber
\EQN
We have omitted  the 6-form current since the supergravity 
fluctuations do not contain it directly. 
The convention is that the indices $i,j, \ldots$ run over 9 spatial dimensions, and the indices $a, b,\ldots$ do over the 
10 dimensions (=time + 9 spatial dimensions).  
The field strength $F_{ab}$ thus consists of 2 part, 
$F_{0i}=\dot{X}_i $ and $F_{ij}=[X_i, X_j]/\ell_s^2$. 
To avoid possible confusion, we remark that the light-cone 
indices on these operators must be interpreted as being for 
the current densities before integration. There is implicitly 
a hidden integration over $x^-$ which is not manifest 
in Matrix theory.    
The generalized conformal dimensions  of  these operators are given in Tables 3 and 4. 

\vspace{1cm}
\begin{center}
{\bf Table 3}
\[
\begin{array}{|c||c|c|c|c|c|c|}
\hline
\mbox{currents} & T^{++}&T^{+i}& T^{+-} & T^{ij} & T^{-i} & T^{--} \\
\hline\hline
\mbox{dimensions}& -3 & -1 & 1 & 1 & 3 &5 \\
\hline
\end{array}
\]

\vspace{1cm}
{\bf Table 4}
\[
\begin{array}{|c||c|c|c|c|}
\hline
\mbox{currents} &J^{+ij}&J^{+-i}& J^{ijk} & J^{-ij}   \\
\hline\hline
\mbox{dimensions}& -1 & 1 & 1 &  3 \\
\hline
\end{array}
\]
\end{center}

\vspace{1cm}
Comparing the tables 3 and 4 with the tables 1 and 2, respectively, 
it is natural to make the identifications listed in tables 5 and 6 
below between the Matrix-theory operators and the supergravity 
fluctuations at the boundary.  

\vspace{1cm}
\begin{center}
{\bf Table 5}
\[
\begin{array}{|c||c|c|c|c|c|c|}
\hline
\mbox{Matrix operators} & T_{\ell}^{++}&T_{\ell}^{+i}& \tilde{T}_{\ell}^{+-} & T_{\ell}^{ij} & T_{\ell}^{-i} & T_{\ell}^{--} \\
\hline\hline
\mbox{SUGRA modes} & s^{\ell}_3 & a^{\ell}_2 &s^{\ell}_2  &  b^{\ell} & a_1^{\ell} &s_1^{\ell} \\
\hline
\end{array}
\]

\vspace{1cm}
{\bf Table 6}
\[
\begin{array}{|c||c|c|c|c|}
\hline
\mbox{Matrix operators} &J_{\ell}^{+ij}&J_{\ell}^{+-i}& J_{\ell}^{ijk} & J_{\ell}^{-ij}   \\
\hline\hline
\mbox{SUGRA modes} &v^{\ell}_2 & u^{\ell} & \alpha^{\ell} &  v_1^{\ell} \\
\hline
\end{array}
\]
\end{center}

\vspace{1cm}

The indices $\ell$ in these tables denote the $\ell$-th component 
in the harmonic expansion.  On the side of Matrix theory, 
the corresponding operators up to possible 
ordering ambiguity are, {\it e. g.}, 
\EQAN
T_{\ell, i_1 i_2 \cdots i_{\ell}}^{++}&=&{1\over R}{\rm STr}(\tilde{X}_{i_1}\tilde{X}_{i_2}
\ldots \tilde{X}_{i_{\ell}} +\cdots ) , \quad  (\ell\ge 2)\\
T_{\ell,i_1 i_2 \cdots i_{\ell}}^{+i}&=&{1\over R}{\rm STr}(\dot{X}_i
\tilde{X}_{i_1}\tilde{X}_{i_2}
\ldots \tilde{X}_{i_{\ell}} +\cdots ) , \quad (\ell\ge 2)\\
&&  \qquad  etc, 
\EQNN
where the coordinate matrices corresponding to the nonzero orbital angular 
momentum are normalized by dividing by the radial distance $q^{1/7}$
of the boundary,  $\tilde{X}_i \equiv X_i/q^{1/7}$ which accounts for the 
coefficient $4/7=1-3/7$ of $\ell$ in the formula of the generalized conformal 
dimensions (\ref{confdimen}).   The noncontracted spatial indices of the currents here only take the 
values from the 8 dimensional space S$^8$ instead of the 
full 9 dimensional space. Note also that the trace part of the 
noncontracted indices should be subtracted, corresponding to 
the 8 dimensional 
harmonic expansion.  For example, the operator 
$T^{ij}_{\ell}$ in the Table 5 must be traceless with respect to the symmetric tensor indices $ij$ and to 
the orbital part, separately.  The tilde on the 
operator $\tilde{T}^{+-}_{\ell}=T^{+-}_{\ell} 
+ c T^{ii}_{\ell}$ 
indicates that this operator can mix \cite{taylor2}
\footnote{We would like to thank W. Taylor for 
a comment on this.}  with the 
trace part of $T^{ij}_{\ell}$ whose coefficient $c$ 
cannot be predicted from our results alone.      

On the supergravity side, the components  which have indices along the radial direction 
  and also the ones with 
 lower angular momentum than the restriction indicated 
in Tables 1 and 2 are either pure gauge modes or not independent physical propagating modes in the bulk.   

The agreement of the generalized conformal dimensions 
between the fluctuations of supergravity and the Matrix-theory 
operators is not surprising if we consider it from 11 dimensional viewpoint of discrete light-cone quantization 
(DLCQ). 
The reason is (see the 
second paper in ref. \cite{li-yo}) that the scaling transformations  
(\ref{oppscaling}) and (\ref{couplingscaling}) are equivalent, 
as was already mentioned in section 2,  
to the following boost-like transformation (Minkowski metric)  
\EQ
t \rightarrow \lambda^{-2} t , \quad R \rightarrow \lambda^2 R .
\quad X_i \rightarrow X_i . 
\label{scale1}
\EN
 Note that we have 
shifted from the original string unit to the new system of unit in 
which the 11 dimensional Planck length 
$g_s^{1/3}\ell_s$ is kept invariant by making the global scaling 
transformation $X_i \rightarrow \lambda^{-1} X_i, 
\quad t\rightarrow \lambda^{-1}t, \quad 
\ell_s \rightarrow \lambda^{-1} \ell_s$.  This leads to 
$q/r^7  \sim \ell_P^9/R^2r^7  \rightarrow \ell_P^9/\lambda^2R^2r^7$. 
 Then the 
11 dimensional metric for small $R$, 
\EQ
ds_{11}^2 = 2dx^+dx^- + {q\over  r^7}dx^-dx^- + dx^idx^i ,
\label{11dmetric}
\EN
($x^-=x_{10}-t, x^+ =(x_{10}+ t)/2$) corresponding to the 
classical D-particle solution is invariant under the boost 
$
x^+ \rightarrow \lambda^{-2} x^+ , \quad R \rightarrow \lambda^2 R .
\quad X_i \rightarrow X_i ,
$
which is almost equivalent to (\ref{scale1}) except 
for the identification of the time variable, 
provided that we interpret  the compactification radius $R$ to be along the 
$x^-$ direction $x^-\sim x^-+2\pi R$. .  Although the reduction to 10 dimensions in general {\it breaks} the 
boost invariance by setting  the dilaton 
to be of the form (\ref{dilaton}),  
the symmetry is recovered in the near-horizon limit. 
In fact, it is easy to check that taking the near horizon limit in 10 dimensions is equivalent to modify the 11 dimensional metric (\ref{11dmetric}) to 
\[
ds_{11}^2 = 2dt  dx^- + {q\over r^7}dx^-dx^- + dx^idx^i ,
\]
which is indeed invariant under (\ref{scale1}). 
Remember that the 10D and 11D metrics are 
related by $ds_{11}^2=e^{-2\tilde{\phi}3}ds_{10}^2 
+e^{4\tilde{\phi}/3}(dx_{10}-A_0 dt)^2$. 
Thus we expect that the 
fluctuations around the D-particle solution in the near 
horizon limit is 
classified by the transformation property corresponding to the 
 boost if we reinterpret the time as the light-cone time 
$t \rightarrow x^+$ which is 
understandable in the limit of small compactification 
radius $R$ along the $x^-$ direction. The dimensions indicated in the tables 3 and 4 agree 
precisely with those expected from 
the boost transformation after making the shift 
corresponding to the change of unit: 
The resulting dimensions are uniformly shifted by one unit 
from the generalized conformal dimensions and are given by the formula
$2(n_--n_+)+2 + 4\ell/7$. The additional factor 2 corresponds to 
the hidden integration over $x^-$. 
  Thus the agreement 
of the dimensions on both sides is just as it should be. 
This provides strong evidence 
for the consistency of Matrix theory 
with the DLCQ interpretation for large $N$, 
conforming to the results of perturbative calculations 
 \cite{bbpt} \cite{okawayoneya} at fixed $N$.   

However,  in general,  the knowledge of the dimensions is 
not sufficient to fix the form of the correlators 
even for 2-and 3-point functions, in contrast 
to the case of usual conformal symmetry.
Once the dependence on the coupling $g_s$ is given, 
we would be able to fix the scaling behavior 
with respect to the time differences.  Namely,  the coupling constant dependence 
and the scaling behavior with respect to time are only simultaneously 
determined.
 We should recall here that in the usual conformal case, the coupling constant 
is invariant under the conformal transformation, so that 
we cannot fix the coupling constant dependence 
of the correlators by conformal symmetry.  
We again need an explicit 
computation to determine it. Thus the strength of the constraint that the 
generalized conformal transformation puts on Green functions 
is not less than the one we had in the case of
the ordinary conformal symmetry.  
For example, even though the Wilson loop in the 
case of AdS$_5\times $S$^5$ exhibits the Coulomb behavior, 
AdS-CFT correspondence predicts \cite{wilsonloop}  a very nontrivial 
behavior $(g_{{\rm YM}}^2N)^{1/2}$,  instead of 
$g_{{\rm YM}}^2N$ of the free theory,  for the effective (charge)$^2$, suggesting the existence of a screening effect 
of factor $1/g_sN$ 
 due to complicated large $N$ dynamics. 
  In a similar sense, 
our result  (\ref{2point}) for the correlators  gives nontrivial 
predictions for the 2-point correlators of Matrix theory 
in the large $N$ limit.  

We remark that in the particular problem treated in the present paper, 
it is actually possible to predict the $g_s$ (and $N$)  dependence of two point correlators from the generalized conformal dimension $\Delta$ as 
\[
g_s^{(\Delta + \Delta_e -3)/5 } = g_s^{{2\over 7}\nu -1 }
\]
where $\Delta_e=-1$ is the engineering 
dimension of the operator, 
using the fact that, apart from the Newton constant $\kappa^2$, 
$g_s$ only appears in the combination 
$q\propto g_sN\ell_s^7$.

   From a purely 10 dimensional viewpoint, we can consider the 
non-extremal black hole solution corresponding to 
D0-branes \cite{pol2}, whose near horizon geometry 
is described by  
\EQ
ds^2 = -e^{-2\tilde{\phi}/3}(1-\bigl({r_0\over r}\bigr)^7) dt^2 + e^{2\tilde{\phi}/3}(1-\bigl({r_0\over r}\bigr)^7)^{-1} dr^2 
+e^{2\tilde{\phi}/3}r^2 d\Omega_8^2 .
\EN
The Hawking temperature and the 
entropy is given, up to numerical coefficients,  by
\EQ
T_H \sim (g_sN)^{-1/2}\Bigl({r_0\over \ell_s}\Bigr)^{5/2}
\ell_s^{-1}, \quad 
S\sim N^2(g_sN)^{-3/5}\Bigl(\ell_sT_H)^{9/5} ,
\EN
 within the range of validity of the 10 dimensional picture 
$1\ll g_sN(T_H\ell_s)^{-3} \ll N^{10/7}$.
The correlator (\ref{tstensorcorrelator}) for the 
traceless symmetric tensor part of the metric perturbation  
gives precisely 
the same $N$ and $g_s$ dependencies in the part which is 
independent of $\ell$.  The agreement  may be regarded as evidence for the fact that the 
correlator corresponding to the energy-momentum tensor 
without mixing of other modes adequately counts the number of degrees of freedom in the low-energy 
regime of many D-particle dynamics.  

Let us finally turn to  the crucial problem, namely, the possible implications  
of our results on the 
Matrix theory conjecture as originally proposed in \cite{bfss}.  
As summarized in section 2, the basic assumption 
behind this conjecture is that 
the infinite-momentum frame (IMF) is achieved by taking 
the large $N$ limit with the compactification radius 
along the 11th spatial direction being kept fixed. 
If the conjecture is valid, the large $N$ behavior must 
also be consistent with the different identification 
of the boost transformation in sending the 
system to IMF.  
Namely, the boost factor is proportional to $N$, 
\EQ
\tau \rightarrow N\tau, \quad P^-\rightarrow {1\over N} P^-, 
\quad P^+\rightarrow N P^+,   \quad \ldots .
\label{nibl}
\EN
Note that now $R$, $\ell_P$ and 
the transverse coordinates $X_i$  are all fixed. 
This  scaling  is indeed consistent with the 
form of the Hamiltonian (\ref{hamiltonian}) as 
already reviewed in section 2. 
This limit is, however,  entirely different from the 
usual `t Hooft limit in which we keep $g_sN\propto g_{{\rm YM}}^2N$ fixed and then  (\ref{2point}) would be 
proportional to $N^2$  as it should be.  
In the  limit  (\ref{nibl}), on the other hand, 
 the 2-point functions behave quite differently 
as 
\EQ
\left<{\cal O}(\tau){\cal O}(\tau ')\right> 
\rightarrow N^{-{12\nu\over 7}} G(\tau-\tau') .
\EN
We parametrize the order $\nu$ of Bessel function as 
\[
\nu={7\over 5}(1-n_++n_- ) + {2\ell\over 5}, 
\]
where $n_{\pm}$ is the number of light-cone indices 
$\pm$ of the corresponding Matrix operators.  
Using this result, we can define the effective 
boost dimensions of the operators  by
\[
d_{IMF} = {6\over 5}(n_+-n_- -1) -({1\over 5}+{1\over 7})\ell .
\]
The factor $1/7$ in the last parenthesis is 
canceled by the normalization factor in the 
harmonic expansion, while the factor $-1 $ in the first 
parenthesis should correspond to the hidden integration over 
$x^-$. 
It is now possible to assign the dimensions 
 $N^{\pm 6/5}=N^{\pm 1}N^{\pm 1/5}$ 
to the upper 
light-cone indices $\pm$ respectively, and $N^{-1/5}$ to each 
orbital factor $X_i$ along $S^8$ in the harmonic  
expansion.  Namely, the dimension $d_{IMF}$ 
is determined solely by the external space-time indices 
of the operator.  This itself is a nontrivial 
phenomenon,  suggesting that the large $N$ limit 
is indeed connected with some space-time symmetry 
of Matrix theory.  
It implies the validity of a large $N$ renormalization group equation of the following type 
\[
\Big[N{\partial \over \partial N} 
+\sum_{i=1}^n \Bigl(\tau_i {\partial \over \partial \tau_i} -d_{IMF, i}
\Bigr)\Big] \langle {\cal O}_1(\tau_1){\cal O}_2(\tau_2) \cdots 
{\cal O}_n(\tau_n) \rangle_c =0
\]
for general $n$-point correlation functions of Matrix theory 
{\it before} making the scaling transformation (\ref{nibl}).  

However, the usual kinematics would require that the 
scaling factor associated to the light-cone indices
 be $N^{\pm 1}$ instead of $N^{\pm 6/5}$.  How to interpret the 
anomalous factor $N^{\pm 1/5}$ is not clear to us. 
Is it correlated with the same power 
factor $N^{-1/5}$ associated to 
the orbital factor along $S^8$? 
In view of holography \cite{bfss}, on the other hand, the latter 
behavior $N^{-1/5}\rightarrow 0$ is quite puzzling at least  
apparently,  
since the transverse size of the system should increase 
as the number of partons increases.  It is not completely 
obvious to us, however, whether  
this implies an immediate contradiction 
with holography.   
It might indicate 
some kind of screening effect in the large $N$ limit 
with respect to the effective 
sizes of the states for higher angular momentum 
as seen from physical operators.  
This is somewhat reminiscent of the behavior 
of the Wilson loop in the case of AdS$_5$-SYM 
correspondence as mentioned above.  
Remember that the increase of transverse size 
under the boost usually associated with holography
is  itself a very puzzling behavior and certain  
screening must be operative for its resolution. 

Let us  examine the range of 
validity of the above predictions on the large $N$ IMF.  
We have emphasized in section 2 that the range of validity of the  generalized AdS-CFT correspondence for Matrix theory is limited by  an infrared 
cutoff of order $(g_sN)^{1/7}\ell_s$ in the large $N$ limit 
for $g_sN>1$.  In the large $N$ limit 
with a small but fixed $g_s$, this cutoff is bigger than the 
transverse extension proportional to $ N^{1/9}$ of the typical states derived  
in mean field approximation \cite{bfss} (or 
$N^{11/81}$ in the Thomas-Fermi approximation \cite{aclm}) in effectivce 
theory for diagonal components, 
but is smaller than the more reliable estimate $N^{1/3}$ obtained by use of  the virial theorem \cite{pol}, 
which explicitly takes account into 
the fluctuations of off-diagonal components.   
Therefore, it seems that the limitation in the 
infrared region is rather serious. This strongly suggests that 
our results should be regarded as predictions for the correlators of Matrix theory put in a `small' box from the point of view 
of the Matrix-theory conjecture.  The anomalous scaling 
behavior may then be interpreted as an artifact caused by 
the finite size effect.  The latter effect may contribute to 
the lessening of the degrees of freedom\footnote{
For a further discussion on this point, see ref. \cite{yostring99}. }. 

On the other hand, in the short distance limit for the 
validity of the generalized AdS-CFT correspondence 
was (\ref{smallcoupling}) , 
$r_{10}\equiv g_s^{1/3}N^{1/7} \ell_s< r$,  in the 10 dimensional picture. For a small but fixed $g_s$ as required 
by the large $N$ IMF, the lower limit increases in the 
same order as the near-horizon limit. 
This certainly makes dubious our procedure in extracting the 
correlators from the near horizon boundary.  
Formally, however, 
the contribution at the inside boundary 
still vanishes exponentially in the large $N$ limit 
for fixed but sufficiently small $g_s$.  
Remember that we have normalized the 
solutions at the outer near-horizon boundary.  
It is also possible to take the limit such that both $g_sN 
\rightarrow \infty$ and vanishing of the ratio 
of two distances,  
$r_{10}/q^{1/7}\rightarrow 0$,  
are valid simultaneously in the limit of large $N$ 
by allowing that $g_s$ slowly changes with large $N$.  
This suggests that our result may be continued to the 
region of the large $N$ IMF at least in the 
short distance region.  
Furthermore, we can improve the situation  by going to the 11 dimensional picture. 
In the latter, the short distance condition for the validity 
of the classical approximation is (\ref{11dcurvature}),  
$r_{11} \equiv g_s^{5/27}N^{1/9}\ell_s < r$. 
Since $r_{11}/q^{1/7}= g_s^{8/189}N^{-2/63}
\rightarrow 0$
 (or $z_{11}/q^{1/7} =g_s^{-20/189}N^{5/63} \rightarrow \infty$) 
 for $N\rightarrow 
\infty$ for fixed $g_s$, 
our procedure of extracting the 
correlators from the near-horizon boundary in this 
picture is safer than in the 
10 dimensional picture. 
 
Do the limitations which we have discussed here explain the above anomalous behavior? 
Or is it related to a different limitation that we are considering 
the small coupling region $R\ll \ell_P$?  Note that 
from the viewpoint of 11 dimensions the parameter 
$R$ only appears as an overall scale factor of the 
Hamiltonian (\ref{Hamilton}). This seems to indicate
 that the scaling  
property might not be affected by the condition $R\ll \ell_P$.  
 In any case, it is necessary 
to clarify these problems before drawing 
definite conclusions on the Matrix theory 
conjecture from our predictions on the correlation functions. 
We hope that our predictions will be useful for 
future investigations.  

\vspace{0.5cm}
\noindent
Acknowledgements

Part of the present work was done while one (TY) of the present authors was visiting
 Stanford University.    
TY would like to thank L. Susskind and S. Shenker for 
hospitality and discussions in a seminar. 
He also thanks A. Jevicki and Y. Kazama for conversations  
and M. Li for e-mail exchanges related to this subject 
at an early stage of the present work. YS thanks T. Muto, T. Kitao and S. Tamura for discussions. 
Finally, the work  is supported in part 
by Grant-in-Aid for Scientific  Research (No. 09640337) 
and Grant-in-Aid for International Scientific Research 
(Joint Research, No. 10044061) from the Ministry of  Education, Science and Culture.


\vspace{1cm} 
\noindent
{\large Appendix}
\renewcommand{\theequation}{A.\arabic{equation}}
\appendix 
\setcounter{equation}{0}
\vspace{0.3cm}
\noindent
Spherical harmonics on $S^N$

\vspace{0.3cm}

 Any function defined on $S^N$ can be expanded into
the set of the irreducible representations of $SO(N+1)$, namely
the spherical harmonics.
We briefly summarize the definitions and the properties
of the spherical harmonics.
Details on the spherical harmonics in general dimensions can be
found {\it e.g.} in \cite{sphharmo}.

A scalar function $\hat{\phi}(r, x^i) $, where 
$r$ is the radius of the sphere $S^N$ and 
$x^i \, \, (i=1, ...,N+1; x^ix^i=1)$ are normalized Cartesian 
coordinates on the sphere, can be expanded as
\EQ
\hat{\phi}=\sum \varphi (r) Y(x^i).
\EN
$Y$ is called the scalar harmonics whose explicit form 
is given by 
\EQ
Y= C_{ m_1\ldots m_\ell}x^{m_1}\cdots x^{m_\ell}\qquad(\ell=0,1,\ldots)
\EN
where $C_{ m_1\ldots m_\ell}$ is totally symmetric and traceless  in its indices. In what follows we set $N=8$. 
We have suppressed the indices which label the harmonics  
here and in the text for the sake of brevity. 
Note that harmonics with a given $\ell$ transform irreducibly 
under $SO(9)$ and
the number of independent harmonics is given by the dimension of
the representation. Spherical harmonics are the eigenfunctions
of the Laplacian on the sphere (which corresponds to the second order
Casimir of SO(9)). The eigenvalue evaluated on the unit sphere $S^8$ is
\EQ
D^i D_i Y =-\ell(\ell+7) Y.
\label{scalarev} 
\EN

A vector function $\hat{A}_i$ on the sphere can be written as a sum of the 
divergenceless part and the derivative of a scalar
\EQ
\hat{A}_i=\sum a(r) Y_i(x^i) +\sum \bar{a}(r) D_i Y(x^i).
\EN
Divergenceless vector $Y_i$ is called the vector harmonics. 
The explicit form is given by
\EQ
Y_n= C_{ n m_1 \ldots m_\ell}x^{m_1}\cdots x^{m_\ell}\qquad(\ell=1,2\ldots).
\EN
 The coefficients $C_{ n m_1 \ldots m_\ell}$  are 
antisymmetric under the exchange of the first two indices $(n,m_1)$, 
totally symmetric and traceless with respect to  $m_1, 
\ldots ,m_\ell$. The first condition is to ensure that the vector 
is tangent to the 
sphere. We need $\ell\ge 1$ to satisfy the conditions.
The eigenvalue of the laplacian is 
\EQ
D^j D_j Y_i =[-\ell(\ell+7)+1] Y_i
\EN 
If we impose the condition $D^i \hat{A}_i =0$, 
the expansion must be
\EQ
\hat{A}_i =\sum a Y_i.
\EN
which is easily verified using (\ref{scalarev}) and $D_i Y^{(\ell=0)}=0$.

Symmetric traceless tensor on $S^8$ is written as the sum of divergenceless 
part,
derivative of a divergenceless vector and second derivative of
a scalar.
\EQ
h_{ij}-g_{ij} h^k_k=\sum b(r) Y_{ij}(x^i)+\sum \bar{b}(r)(D_iY_j+D_jY_i)(x^i)
+\sum \bar{\bar{b}}(r)(D_iD_j-g_{ij}D^kD_k)Y(x^i)
\EN
The explicit form of the tensor harmonics $Y_{ij}$ is given by
\EQ
Y_{n_1 n_2}=C_{ n_1 n_2 m_1 \ldots m_\ell}x^{m_1}\cdots x^{m_\ell}\qquad(\ell=2,3\ldots).
\EN
 The coefficients $C_{ n_1 n_2 m_1 \ldots m_\ell}$ must be 
antisymmetric under the exchange of $(n_1,m_1)$ or $(n_2,m_2)$ to ensure
that the tensor indices are tangent to the sphere, 
symmetric under the exchange of $(n_1,n_2)$
to make the tensor indices symmetric,
and totally symmetric and traceless  
with respect to $m_1 \ldots m_\ell$. We need $\ell\ge 2$ to meet the conditions.
The eigenvalue of the Laplacian is 
\EQ
D^k D_k Y_{ij} =[-\ell(\ell+7)+2] Y_{ij}. 
\EN
If we impose the condition $D^i (h_{ij}-g_{ij} h^k_k)=0$, it can 
be proved that the expansion is reduced to 
\EQ
h_{ij}-g_{ij} h^k_k=\sum b Y_{ij}.
\EN

Similarly,  $p$-form $\hat{A}_{i_1\ldots i_p}$ $(p=1,2,3)$ 
can be written as the sum of
the divergenceless part and the exterior derivative of $(p-1)$-form
\EQ
\hat{A}_{i_1\ldots i_p}(x^i)=\sum \alpha Y_{[i_1\ldots i_p]}(x^i)
+\sum \bar
{\alpha}
D_{[i_1} Y_{[i_2\ldots i_p]]}(x^i) 
\EN
The explicit form of the $p$-form harmonics $Y_{[i_1\ldots i_p]}$ is
\EQ
Y_{[n_1\ldots n_p]}=C_{ n_1 \ldots n_p m_1 \ldots m_\ell}x^{m_1}\cdots x^{m_\ell}\qquad(\ell=1,2\ldots).
\EN
The coefficients $C_{ n_1 \ldots n_p m_1 \ldots m_\ell}$ are 
antisymmetric under the interchange of $m_1$ with any one of
$n_1,\ldots,n_\ell$ and totally symmetric and traceless 
with respect to$(m_1 \ldots m_\ell)$.
To meet the conditions, we need $\ell\ge 1$.
Under the condition $D^{i_1} \hat{A}_{i_1\ldots i_p}=0$,
the expansion becomes
\EQ
\hat{A}_{i_1\ldots i_p}=\sum \alpha Y_{[i_1\ldots i_p]}
\EN
The eigenvalue of the Laplacian is
\EQ
D^k D_k Y_{[i_1\ldots i_p]} =[-\ell(\ell+7)+p] Y_{[i_1\ldots i_p]}.
\EN

\end{document}